\documentclass[structabstract]{aa}
\usepackage{txfonts}
\usepackage{natbib}
\usepackage{graphicx}
\usepackage{aalongtable}
\bibpunct{(}{)}{;}{a}{}{,}

\begin{document}

\title{Two new SB2 binaries with main sequence B-type pulsators in the \textit{Kepler} field.\thanks{Based on observations made with the Mercator telescope, operated by the Flemish Community, with the Nordic Optical Telescope, operated jointly by Denmark, Finland, Iceland, Norway, and Sweden, and with the William Herschel Telescope operated by the Isaac Newton Group, all on the island of La Palma at the Spanish Observatorio del Roque de los Muchachos of the Instituto de Astrof\'{i}sica de Canarias.}\fnmsep\thanks{Based on observations obtained with the \textsc{Hermes} spectrograph, which is supported by the Fund for Scientific Research of Flanders (FWO), Belgium, the Research Council of KU Leuven, Belgium, the Fonds National Recherches Scientific (FNRS), Belgium, the Royal Observatory of Belgium, the Observatoire de Gen\`{e}ve, Switzerland and the Th\"{u}ringer Landessternwarte Tautenburg, Germany.}}

\author{P.~I.~P\'{a}pics\inst{\ref{inst1}}
\and A.~Tkachenko\inst{\ref{inst1}}\thanks{Postdoctoral Fellow of the Fund for Scientific Research (FWO), Flanders, Belgium} %spectral disentangling and abundances
\and C.~Aerts\inst{\ref{inst1},\ref{inst2}}
\and M.~Briquet\inst{\ref{inst4}}\thanks{F.R.S.-FNRS Postdoctoral Researcher, Belgium.}
\and P.~Marcos-Arenal\inst{\ref{inst1}} %12 spectra
\and P.~G.~Beck\inst{\ref{inst1}} %7 spectra
\and K.~Uytterhoeven\inst{\ref{inst5},\ref{inst6}} %7 spectra (PI)
\and A.~Trivi\~{n}o~Hage\inst{\ref{inst5},\ref{inst6}} %7 spectra
\and J.~Southworth\inst{\ref{inst3}} %4 spectra
\and K.~I.~Clubb\inst{\ref{inst7}} %3 spectra
\and S.~Bloemen\inst{\ref{inst1}} %1 spectra + pixelextraction
\and P.~Degroote\inst{\ref{inst1}} %distortion, inclination checks
\and J.~Jackiewicz\inst{\ref{inst8}} %2 spectra
\and J.~McKeever\inst{\ref{inst8}} %2 spectra
\and H.~Van~Winckel\inst{\ref{inst1}} %1 spectra
\and E.~Niemczura\inst{\ref{inst11}} %1 spectra (PI)
\and J.~F.~Gameiro\inst{\ref{inst9},\ref{inst10}} %1 spectra
\and J.~Debosscher\inst{\ref{inst1}}} %sample selection

\institute{Instituut voor Sterrenkunde, KU Leuven, Celestijnenlaan 200D, B-3001 Leuven, Belgium \email{Peter.Papics@ster.kuleuven.be}\label{inst1}
\and Department of Astrophysics, IMAPP, University of Nijmegen, PO Box 9010, 6500 GL Nijmegen, The Netherlands\label{inst2}
\and Institut d'Astrophysique et de G\'{e}ophysique, Universit\'{e} de Li\`{e}ge, All\'{e}e du 6 Ao\^{u}t 17, B\^{a}t B5c, 4000 Li\`{e}ge, Belgium\label{inst4}
\and Instituto de Astrofisica de Canarias, 38205 La Laguna, Tenerife, Spain\label{inst5}
\and Dept. Astrof\'{i}sica, Universidad de La Laguna (ULL), Tenerife, Spain\label{inst6}
\and Astrophysics Group, Keele University, Staffordshire, ST5 5BG, United Kingdom\label{inst3}
\and Department of Astronomy, University of California, Berkeley, CA 94720-3411, USA\label{inst7}
\and Department of Astronomy, New Mexico State University, Las Cruces, NM 88001, USA\label{inst8}
\and Instytut Astronomiczny, Uniwersytet Wroc\l{}awski, Kopernika 11, 51-622 Wroc\l{}aw, Poland\label{inst11}
\and Centro de Astrof\'{i}sica, Universidade do Porto, Rua das Estrelas, 4150-762 Porto, Portugal\label{inst9}
\and Departamento F\'{i}sica e Astronomia, Faculdade de Ci\^{e}ncias da Universidade do Porto, Portugal\label{inst10}}

\date{Received ? ???? 2013 / Accepted ? ???? 2013}

\abstract{OB stars are important in the chemistry and evolution of the Universe, but the sample of targets well understood from an asteroseismological point of view is still too limited to provide feedback on the current evolutionary models. Our study extends this sample with two spectroscopic binary systems.}{Our goal is to provide orbital solutions, fundamental parameters and abundances from disentangled high-resolution high signal-to-noise spectra, as well as to analyse and interpret the variations in the \textit{Kepler} light curve of these carefully selected targets. This way we continue our efforts to map the instability strips of $\beta$\,Cep and slowly pulsating B stars using the combination of high-resolution ground-based spectroscopy and uninterrupted space-based photometry.}{We fit Keplerian orbits to radial velocities measured from selected absorption lines of high-resolution spectroscopy using synthetic composite spectra to obtain orbital solutions. We use revised masks to obtain optimal light curves from the original pixel-data from the \textit{Kepler} satellite, which provided better long term stability compared to the pipeline processed light curves. We use various time-series analysis tools to explore and describe the nature of variations present in the light curve.}{We find two eccentric double-lined spectroscopic binary systems containing a total of three main sequence B-type stars (and one F-type component) of which at least one in each system exhibits light variations. The light curve analysis (combined with spectroscopy) of the system of two B stars points towards the presence of tidally excited $g$ modes in the primary component. We interpret the variations seen in the second system as classical $g$ mode pulsations driven by the $\kappa$ mechanism in the B type primary, and explain the unexpected power in the $p$ mode region as a result of nonlinear resonant mode excitation.}{}

\keywords{Asteroseismology - 
Stars: variables: general - 
Stars: abundances -
Stars: oscillations - 
Stars: early-type - 
Stars: binaries: general}

\titlerunning{Two new binaries with main sequence B-type pulsators from \textit{Kepler}.}
\maketitle

%%%%%%%%%%%%%%%%%%
%%%Introduction%%%
%%%%%%%%%%%%%%%%%%

\section{Introduction}\label{intro}
In the last 10 years asteroseismology went through an unprecedented evolution. The launch of the MOST \citep[Microvariablity and Oscillations of Stars,][]{2003PASP..115.1023W} mission in 2003 and the CoRoT \citep[Convection Rotation and planetary Transits,][]{2009A&A...506..411A} satellite in 2006 meant the start and the culmination of the industrial revolution of asteroseismology. Shortly afterwards the \textit{belle \'{e}poque} arrived with the launch of the \textit{Kepler} \citep{2010ApJ...713L..79K} satellite in 2009. This immense growth in observational data provides ever better input and testbeds to confront with theoretical calculations of stellar structure and evolution.

Although the primary science-goal of \textit{Kepler} is the detection of Earth-like exoplanets \citep{2010Sci...327..977B}, its virtually uninterrupted photometry also provides us with micromagnitude precision light curves of tens of thousands of stars situated in the 105 square degree field of view (FOV) fixed in between the constellations of Cygnus and Lyra -- a real goldmine for asteroseismology too \citep{2010PASP..122..131G}. While the photometric precision is in the order of the one of CoRoT (despite the fainter apparent magnitude of the typical asteroseismology sample of the \textit{Kepler} mission), the fixed FOV and target list provides a timebase of several years for continuous observations. Such long timebase is necessary to resolve rotationally split multiplets -- which can hardly be achieved from CoRoT data with a time base of 5 months -- and to measure frequencies with the required high precision. The maximal frequency resolution we can aim for is given by the \citet{1978Ap&SS..56..285L} criterion of $\sim2.5/T$ (where $T$ is the timespan of the observations). This yields a resolution of $0.002\,d^{-1}$ in the periodogram, assuming 3.5 years (the originally planned lifespan of the now extended mission) of observations. This resolution of $\sim10^{-3}\,d^{-1}$ is necessary for the secure detection of the rotational splitting of individual modes because of the merging of various multiplets in the typically dense frequency spectra.

OB stars  -- intermediate mass to massive stars -- are important in the chemistry and evolution of the universe. They have a significant contribution to the enrichment of the interstellar matter with heavy elements, moreover, some of them in suitable binary systems can end their lives as Ia type supernovae, the standard candles upon which the study of the expansion of the universe is based.

The first overview on the variability of B-type stars observed by \textit{Kepler} was given by \citet{2011MNRAS.413.2403B}. The authors analysed a selection of 48 main sequence B-type stars, and concluded that 15 of them are slowly pulsating B (SPB) stars (with 7 SPB/$\beta$\,Cep hybrids). Apart from the pulsating stars, they have identified stars with frequency groupings similar to what is seen in Be stars, and suggested that this might be connected to rotation. Some of the stars showed clear sign of modulation due to proximity effects in binary systems or rotation. \citet{2011MNRAS.413.2403B} have found more non-pulsating stars within the $\beta$\,Cep instability strip, and no pulsating stars between the cool edge of the SPB and the hot edge of the $\delta$\,Sct instability strip.

The richness of the observed variable behaviour of B-type stars must imply that details in the internal physics of the various stars, such as their internal rotation, mixing, and convection, etc., are different. Some excitation problems may be solved by increasing the opacity in the excitation layers. Also, binary interaction can complicate the observational data and its interpretation, but possibly provide better constraints than for single stars. In an attempt to increase the number of well studied early-type stars, a sample of B-type stars (having no overlap with the sample analysed by \citeauthor{2011MNRAS.413.2403B} -- thanks to the different selection criteria) was selected for in depth analysis from the \textit{Kepler} field.

%%%%%%%%%%%%%%%%%%
%%%Sample selec%%%
%%%%%%%%%%%%%%%%%%

\section{Target selection}\label{selection}
We have selected targets to construct a new B-type pulsator sample via a very careful iterative process, incorporating several different techniques. As a first step, based on the publicly available Q1 light curves of the \textit{Kepler} mission (release: 15 June 2010), we carried out an automated supervised classification \citep{2011A&A...529A..89D}. This method uses limited Fourier-decomposition of the light curves with a maximum of three dominant frequencies, then after a least-squares fitting, the resulting parameters (frequencies and amplitudes) are compared to the typical values of known groups of pulsating variable stars \citep[for further details -- on the same method applied to the CoRoT exoplanet sample -- see][]{2009A&A...506..519D}. Because the oscillation properties of different classes of pulsators can be very similar, 2MASS colours were used to distinguish between these groups (SPB and $\gamma$\,Dor stars in this case). The remaining nine SPB candidates provided the starting point of our study. None of these pulsators are included in the asteroseismology target list of the mission \citep{2011MNRAS.413.2403B}. The $\gamma$\,Dor stars are analysed by \citet{Tkachenko2013}.

To make sure that not only their oscillation patterns but also their stellar parameters confirm that these stars are indeed B-type main-sequence stars (thus confirming the reliability of the previously outlined classification process), we took high resolution ($R=85\,000$) spectra of the four brightest targets with the \textsc{Hermes} spectrograph \citep{2011A&A...526A..69R} installed on the 1.2 metre Mercator telescope on La Palma, and estimated the fundamental parameters. We carried out a full grid search with four free parameters ($T_\mathrm{eff}$, $\log g$, and  $v \sin i$, plus the metallicity $Z$), using the BSTAR2006 \citep{2007ApJS..169...83L} and ATLAS \citep{2010yCat....102030P} atmosphere grids. The derived parameters place all stars within the SPB instability strip, which already suggested that the rest of the sample -- chosen using the same criteria -- should also fall into that area of the Kiel-diagram. The full spectroscopic campaign (see Sect.\,\ref{spectroscopy}) revealed that two stars are double-lined spectroscopic binaries (SB2), and that one of the fainter objects is not a B star. As an orbital solution can provide additional constraints on the physical parameters of the system, we decided to analyse the two binary targets with first priority. This study is presented in the current paper.

Main sequence B stars are unlikely to have a planetary system, as fierce stellar radiation and winds destroy their primordial circumstellar disk before planets can form, which means that these targets are not interesting for the exoplanet community. As our objects are also not under investigation by the Kepler Asteroseismic Science Consortium \citep[KASC,][]{2010PASP..122..131G}, we had to take special measures to make sure that none of the stars are getting dropped from the target list of \textit{Kepler} during the later phases of the mission. We have successfully applied for \textit{Kepler} Guest Observer (GO) time in Cycles 3 and 4 for additional observations of our targets.

\subsection{A priori knowledge}\label{literature}

\begin{table}
\caption{Basic observational properties of KIC\,4931738 and KIC\,6352430.}
\label{apriori}
\centering
\renewcommand{\arraystretch}{1.25}
\setlength{\tabcolsep}{0pt}
\begin{tabular}{l@{\hskip 12pt} r r@{\hskip 1pt} c@{\hskip 1pt} l r r@{\hskip 1pt} c@{\hskip 1pt} l@{\hskip 6pt} r}
\hline\hline
Parameter && 			\multicolumn{3}{l}{KIC\,4931738} & & \multicolumn{3}{l@{\hskip 6pt}}{KIC\,6352430}& Ref.\\
\hline
$\alpha_{2000}$&&		\multicolumn{3}{l@{\hskip 12pt}}{$\mathrm{ }19^\mathrm{h}36^\mathrm{m}00\fs887$}&&\multicolumn{3}{l@{\hskip 6pt}}{$\mathrm{ }19^\mathrm{h}10^\mathrm{m}57\fs761$}&1\\
$\delta_{2000}$&		$+$&\multicolumn{3}{l@{\hskip 12pt}}{$40\degr05\arcmin55\farcs19$}&$+$&\multicolumn{3}{l@{\hskip 6pt}}{$41\degr46\arcmin33\farcs44$}&1\\
Tycho ID&&				\multicolumn{3}{l@{\hskip 12pt}}{3139-291-1}&&\multicolumn{3}{l@{\hskip 6pt}}{3129-784-1}&1\\
2MASS ID&&				\multicolumn{3}{l@{\hskip 12pt}}{J19360088+4005551}&&\multicolumn{3}{l@{\hskip 6pt}}{J19105776+4146335}&2\\
$B_\mathrm{T}$&&			$11.543$&$\pm$&$0.104$&&$7.859$&$\pm$&$0.008$&1\\
$V_\mathrm{T}$&&			$11.317$&$\pm$&$0.140$&&$7.906$&$\pm$&$0.010$&1\\
$J_\mathrm{2MASS}$&&		$11.449$&$\pm$&$0.021$&&$7.914$&$\pm$&$0.032$&2\\
$H_\mathrm{2MASS}$&&		$11.526$&$\pm$&$0.019$&&$7.936$&$\pm$&$0.016$&2\\
$K_\mathrm{2MASS}$&&		$11.548$&$\pm$&$0.020$&&$7.896$&$\pm$&$0.023$&2\\
\textit{Kepler} mag.&&	$11.645$&&&&$7.958$&&&3\\
\hline
\end{tabular}
\tablebib{(1) \citet{1998A&A...335L..65H}; (2) \citet{2mass}; (3) \citet{2009yCat.5133....0K}.}
\end{table}

\paragraph{KIC\,4931738:} Our first target is a faint star in the constellation of Cygnus. It was also identified -- independent from our study -- as a B-type pulsator by \citet{2012AJ....143..101M}, but no in-depth analysis was carried out, nor was its SB2 nature discovered due to the lack of spectroscopy. The known astrometric and photometric parameters are listed in Table\,\ref{apriori}.

\paragraph{KIC\,6352430:} Our second target (also known as HD\,179506) is -- by \textit{Kepler} standards -- a bright star in the constellation of Lyra at a distance of $495^{+195}_{-109}$ pc \citep[from a parallax of $2.02\pm0.57\,\mathrm{mas}$,][]{2007A&A...474..653V}. It was first described by Sir John Frederick William Herschel in the first half of the 19th century as a visual multiple system (HJ\,2857). The visual components HJ\,2857\,B and HJ\,2857\,C ($V_\mathrm{mag} = 13.6$ and $12.1$) are at a position angle of $\theta = 178\degr$ and $211\degr$ at a distance of $\rho = 14.4\arcsec$ and $36.6\arcsec$, respectively \citep{2000A&A...363..991D}. We show below that the main component itself is a spectroscopic binary (HJ\,2857\,A = KIC\,6352430\,AB). The star was first listed among the $\gamma$\,Dor pulsators by \citet{2011MNRAS.415.3531B}, then \citet{2011MNRAS.415.1691B} points out the large discrepancy between the effective temperature of 9438\,K in the Kepler Input Catalog \citep{2009yCat.5133....0K} and the spectral type of B8\,V \citep{1977MNRAS.180..691H}\footnote{\citeauthor{1977MNRAS.180..691H} note that there was a slight discrepancy between the spectral classification and their $UBV$ photometry - now we know this must have been a sign of binary nature.}, and -- given that the spectral classification as the KIC temperatures are well known to be unreliable in the B star range \citep{2011MNRAS.413.2403B} -- attributes the variation to SPB pulsations. No detailed analysis has been carried out so far for this target either. The known astrometric and photometric parameters are listed in Table\,\ref{apriori}.

%%%%%%%%%%%%%%%%%%
%%%Spectroscopy%%%
%%%%%%%%%%%%%%%%%%

\section{Spectroscopy}\label{spectroscopy}
\subsection{The spectroscopic data}
In order to unravel the binary nature of the two pulsating systems, we have gathered 40 high-resolution spectra of KIC\,4931738 using the \textsc{Hermes} spectrograph \citep{2011A&A...526A..69R} installed on the 1.2 metre Mercator telescope on La Palma (Spain), the \textsc{Fies} spectrograph installed on the 2.5 metre Nordic Optical Telescope on La Palma (Spain), the \textsc{Hamilton} spectrograph \citep{1987PASP...99.1214V} mounted on the 3 metre Shane telescope at the Lick Observatory (USA), the \textsc{Arces} spectrograph \citep{2003SPIE.4841.1145W} mounted on the 3.5 metre ARC telescope at the Apache Point Observatory (USA), and the \textsc{Isis} spectrograph mounted on the 4.2 metre William Herschel Telescope on La Palma (Spain), and 27 high-resolution spectra of KIC\,6352430 using exclusively \textsc{Hermes}. We refer to Table\,\ref{specsummary1} and Table\,\ref{specsummary2} for a summary of the observations.

\subsection{Data reduction}\label{spectralreduction}
We worked with data produced by the standard pipelines of each instrument, except for \textsc{Isis}, where the data were reduced using standard STARLINK routines. We merged the separate orders (taking into account the variance levels in the overlapping range) where it was not yet done or where we felt more comfortable using the separate orders and apply our own merging method rather than using the merged spectra from the pipelines (\textsc{Arces}, \textsc{Fies}, and \textsc{Hamilton} spectra). Subsequently, we performed a careful one-by-one rectification of all spectra -- in the useful range between $4040\AA$ and $6800\AA$. The rectification was done with cubic splines which were fitted through some tens of points at fixed wavelengths, where the continuum was known to be free of spectral lines. For this we utilised an interactive graphical user interface (GUI) which displays a synthetic spectrum (built using a predefined parameter set) in the background enabling the user to select line-free regions. We used these rectified data sets in our further analysis. We note that in case of spectra from the \textsc{Hamilton} spectrograph the pipeline-reduced orders were not blaze-corrected, and the blaze function was not available, so we used a 2D median filter followed by a 2D gaussian filter on the 2D array containing the separate orders to construct an artificial blaze function which we used to correct the spectra. When these filters are parameterised properly -- using suitable widths and sigma values for both dimensions -- the output is well rectified spectral orders. This is a fast and efficient method for the rectification of spectra of early-type stars (whose spectra are dominated by continuum) obtained with \'{e}chelle spectrographs. Furthermore, for the \textsc{Arces} spectra the wings of the Balmer-lines were badly affected while applying the standard blaze-functions, so we could not rectify the region of the Balmer lines properly, but as we did not use them later on, we did not pay any further attention to this issue.

We have placed all observations in the same time frame by converting UTC times of the mid-exposures into Barycentric Julian Dates in Barycentric Dynamical Time (BJD\_TDB, but we simply use the BJD notation from here on), and also calculated the barycentric radial velocity corrections. We checked and corrected for the zero point offset of different wavelength-calibrations by fitting Gaussian functions to a series of telluric lines between $6881\AA$ and $6899\AA$, using an automated normalisation and fitting routine, choosing the first \textsc{Hermes} exposure of KIC\,4931738 as a reference point. The fitted Gaussians also provided the instrumental broadening values for each exposure which we used later in Sect.\,\ref{rvmeasurement}. For the \textsc{Isis} spectra this was not possible because of the wavelength coverage of the spectrograph -- more specifically of the red arm. Here we used a less pronounced region between $6275\AA$ and $6312\AA$ and cross-correlated it with a template spectrum (a slightly broadened version -- to compensate for the different instrumental broadenings -- of the first \textsc{Hermes} exposure) to determine the offset. As the blue arm does not contain any telluric features at all, we had to use an estimation of the instrumental broadening, and to assume that the zero point shift for the two arms is the same. For this reason, we decided not to use radial velocity measurements from this part of the spectrum.

\begin{table*}
\caption{Logbook of the spectroscopic observations of KIC\,4931738 -- grouped by observing runs.}
\label{specsummary1}
\centering
\renewcommand{\arraystretch}{1.1}
\begin{tabular}{c c c c c c c c c}
\hline\hline
Instrument & N &BJD first & BJD last & $\langle\mathrm{SNR}\rangle$ & SNR$_\mathrm{i}$ & $\mathrm{T}_{\mathrm{exp}}$ & R & Observer (PI)\\
\hline
\textsc{Arces}    & 2 & 2456058.93319 & 2456058.95585    & 102 & [100, 103]&  1800         &$31\,500$              & JJ, JMK\\
\textsc{Isis}     & 1 &\multicolumn{2}{c}{2456087.72276} & 101 & [88, 113] &   700         &$22\,000\,\&\,13\,750$\tablefootmark{a} & SB (JS)\\
\textsc{Hamilton} & 3 & 2456132.77847 & 2456134.76473    &  53 & [47,  57] &  1800         &$60\,000$              & JS, KIC\\
\textsc{Fies}     & 7 & 2456146.45631 & 2456148.62473    &  73 & [29,  84] & [391, 2400]   &$25\,000$              & ATH (KU)\\
\textsc{Hermes}   &19 & 2456156.51739 & 2456166.57275    &  32 & [29,  36] & [2700, 3400]  &$85\,000$              & PIP\\
\textsc{Hermes}   & 6 & 2456173.48004 & 2456183.38340    &  32 & [30,  33] & [2200, 2800]  &$85\,000$              & PMA\\
\textsc{Hermes}   & 2 & 2456195.50221 & 2456200.52468    &  31 & [29,  34] &  2700         &$85\,000$              & PB\\
\hline
\end{tabular}
\tablefoot{For each observing run, the instrument, the number of spectra N, the BJD of the first and last exposure, the average signal-to-noise level, the range of SNR values (where the SNR of one spectrum is calculated as the average of the SNR measurements in the line free regions of $4185\AA$ to $4225\AA$, $4680\AA$ to $4710\AA$, $5110\AA$ to $5140\AA$, $5815\AA$ to $5850\AA$, $6180\AA$ to $6220\AA$, and $6630\AA$ to $6660\AA$), the typical exposure times (in seconds), the resolving power of the spectrograph, and the initials of the observer (and the PI if different) are given. \tablefoottext{a}{The resolution for the blue and red arms of \textsc{Isis} is different.}}
\end{table*}

\begin{table*}
\caption{Logbook of the spectroscopic observations of KIC\,6352430 -- grouped by observing runs.}
\label{specsummary2}
\centering
\renewcommand{\arraystretch}{1.1}
\begin{tabular}{c c c c c c c c c}
\hline\hline
Instrument & N &BJD first & BJD last & $\langle\mathrm{SNR}\rangle$ & SNR$_\mathrm{i}$ & $\mathrm{T}_{\mathrm{exp}}$ & R & Observer (PI)\\
\hline
\textsc{Hermes}   & 2 & 2455486.34475 & 2455486.35574       & 118 & [116, 121]&  900        &$85\,000$ & PIP\\
\textsc{Hermes}   & 1 & \multicolumn{2}{c}{2455823.36835}   & 111 &  111      &  900        &$85\,000$ & JFG (EN)\\
\textsc{Hermes}   & 1 & \multicolumn{2}{c}{2456085.53522}   & 134 &  134      & 1000        &$85\,000$ & HVW\\
\textsc{Hermes}   &11 & 2456156.49370 & 2456166.40992       & 113 &  [94, 125]& [900,1200]  &$85\,000$ & PIP\\
\textsc{Hermes}   & 6 & 2456173.50341 & 2456183.34817       & 115 & [100, 122]& [900,1000]  &$85\,000$ & PMA\\
\textsc{Hermes}   & 5 & 2456196.52425 & 2456203.36639       & 105 &  [95, 113]&  900        &$85\,000$ & PB\\
\textsc{Hermes}   & 1 & \multicolumn{2}{c}{2456209.33827}   & 116 &  116      &  900        &$85\,000$ & PIP\\
\hline
\end{tabular}
\tablefoot{Same as for Table\,\ref{specsummary1} but with line free regions of $4680\AA$ to $4700\AA$, $5105\AA$ to $5125\AA$, $5540\AA$ to $5560\AA$, $5820\AA$ to $5850\AA$, $6195\AA$ to $6225\AA$, and $6630\AA$ to $6660\AA$.}
\end{table*}

\subsection{Radial velocities}\label{rvmeasurement}
Radial velocities were measured using several selected narrow lines or line regions by fitting synthetic composite binary spectra to the observed ones. 

First an initial set of fundamental parameters were obtained from a multi-dimensional grid-search ($T_\mathrm{eff}$, $\log g$, $v \sin i$, and the metallicity $Z$, each for both components) using the best (having the highest signal-to-noise ratio (SNR) combined with a good merging and normalisation; for KIC\,4931738 a \textsc{Fies} exposure with an SNR of 84, while for KIC\,6352430 a \textsc{Hermes} exposure with an SNR of 134) available exposure \citep[for a more detailed description of the method see, e.g.,][]{2011A&A...528A.123P}. During the $\chi^2$ calculation we computed the $\chi^2$ values for several line regions (the ones which we later used during the radial velocity measurements, plus $\mathrm{H}_\beta$ and $\mathrm{H}_\gamma$), then summed up the individual $\chi^2$ values and corrected the sum for the number of used line regions. This way narrower regions had the same weight in the final result as broader ones. We used only these selected regions and not the full available wavelength range because we wanted the final result to fit the lines needed for the radial velocity determination the best. We will come back to the derivation of more precise fundamental parameters and abundances in Sect.\,\ref{fundparsandabundances}.

\begin{figure}
\resizebox{\hsize}{!}{\includegraphics{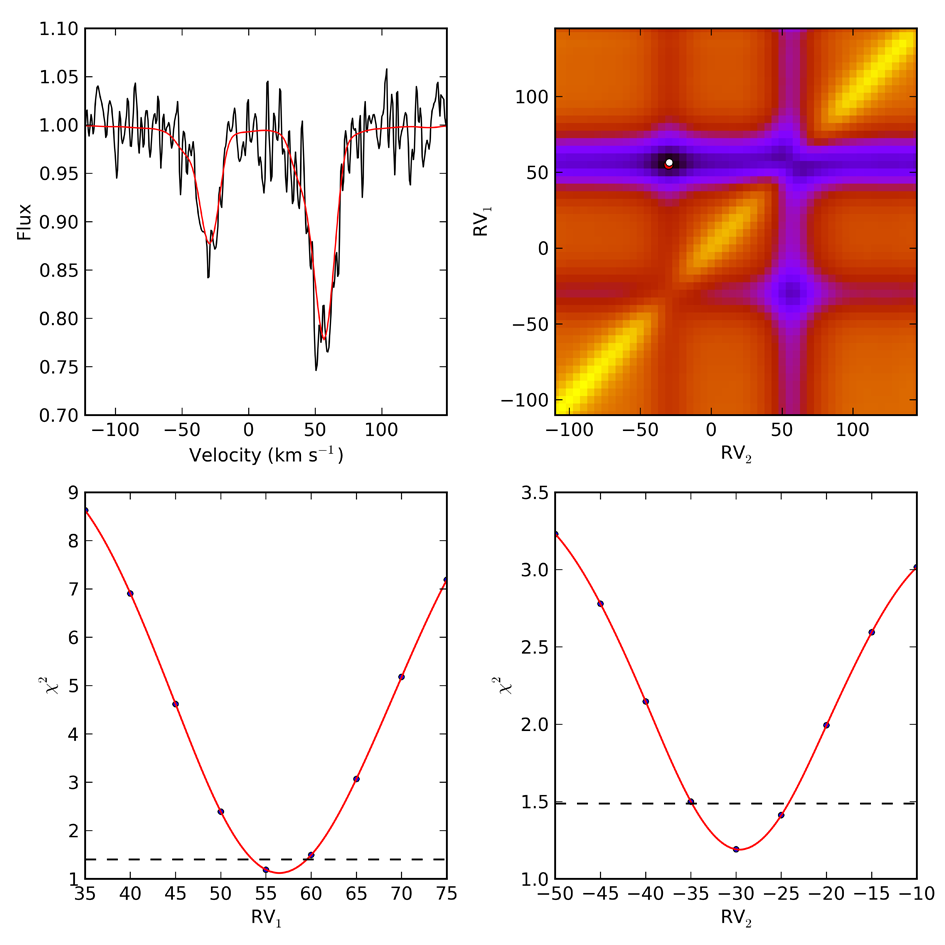}} 
\caption{Visualisation of the radial velocity determination process for KIC\,4931738. (\textit{upper left}) One normalised \textsc{Hermes} spectrum (solid black line) is shown around the \ion{Si}{ii} line at $6347\AA$ with a synthetic binary spectrum (red solid curve) constructed using the determined radial velocities. (\textit{upper right}) The $\chi^2$-space resulting from a full 2D grid search of the radial velocities of the components. The lowest $\chi^2$ value is marked with a red filled circle while the accepted value from fitting the $\chi^2$ values in both dimensions across the best fit value is marked by a white filled circle (almost fully covering the underlying red one). These fits are shown in the (\textit{lower panels}) with a red line, while the $1.25\times\min\{\chi^2_{rv_1,rv_2}\}$ level is plotted with a dashed line. RV$_i$ is expressed in $\mathrm{km\,s}^{-1}$.}
\label{rvchisquare1}
\end{figure}

\begin{figure}
\resizebox{\hsize}{!}{\includegraphics{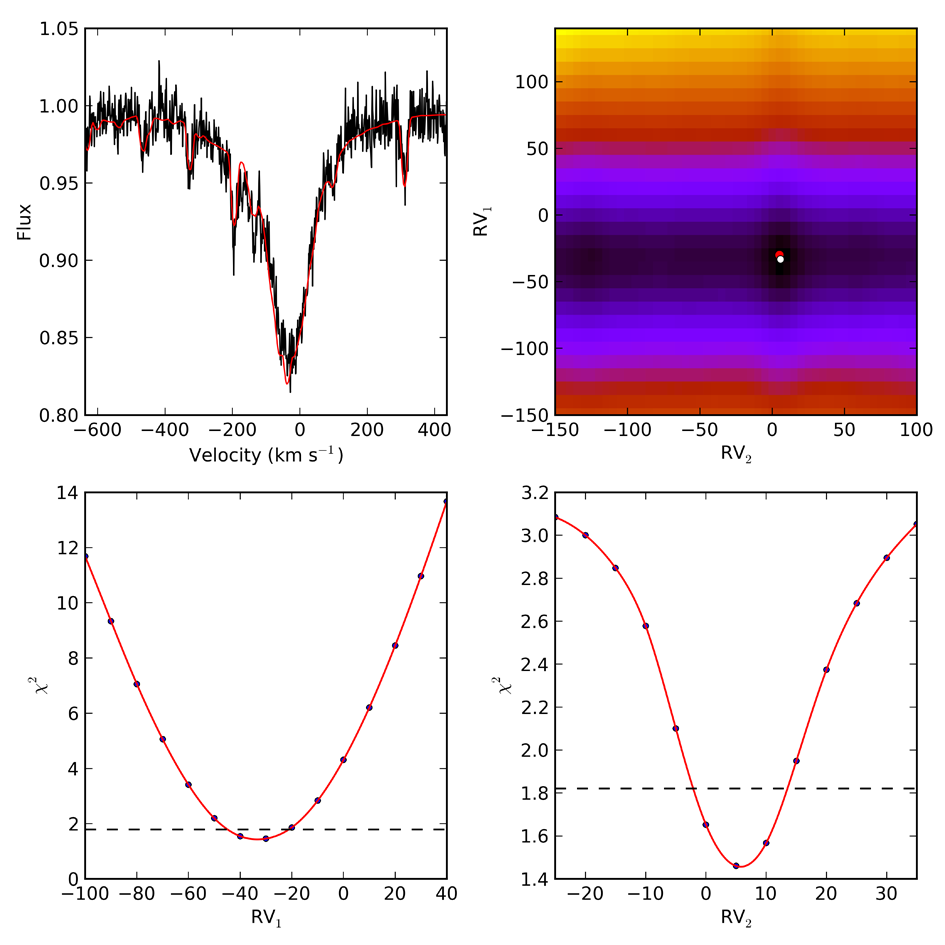}} 
\caption{Same as Fig.\,\ref{rvchisquare1}, but for KIC\,6352430 and the \ion{He}{i} line at $4471\AA$.}
\label{rvchisquare2}
\end{figure}

In the next step we used the derived estimated parameters (see the third sections of Table\,\ref{fundparams1} and Table\,\ref{fundparams2}) to create synthetic spectra of the binary components and fit all observed spectra leaving only the radial velocity values of the components as variables. A full 2D grid search with sufficient resolution was carried out for all line regions independently for all spectra (using specific instrumental broadening values for each exposure derived in Sect.\,\ref{spectralreduction}). In the case of KIC\,4931738 we used the \ion{Mg}{ii} line at $4481\AA$, the \ion{He}{i} lines at $4922\AA$ and $5016\AA$, and the \ion{Si}{ii} lines at $5041\AA$, $5056\AA$, $6347\AA$, and $6371\AA$, while in case of KIC\,6352430 we used the region around the \ion{Fe}{ii} lines at $4550\AA$, $5018\AA$, and $5317\AA$, around the \ion{He}{i} lines at $4471\AA$, $4922\AA$, and $5876\AA$ (this only for the fundamental parameters, not for radial velocities), around the \ion{Mg}{ii} line at $4481\AA$, and around the \ion{Si}{ii} lines at $5041\AA$, $5056\AA$, $6347\AA$, and $6371\AA$ for the measurements, containing both the mentioned lines of the primary, and several additional narrow -- mostly Fe lines -- from the secondary. The best fit values were provided by a third order spline-fit of the $\chi^2$-values in both dimensions, while the error bars were empirically fixed to the radial velocities where $\chi^2_{rv_1,rv_2}=1.25\times\min\{\chi^2_{rv_1,rv_2}\}$. The process is visualised in Fig.\,\ref{rvchisquare1} and Fig.\,\ref{rvchisquare2}. All resulting values were corrected with the barycentric velocities and the zero point offset before proceeding further.

\subsection{Orbital solutions}
The orbital parameters were calculated by fitting a Keplerian orbit through the radial velocity measurements while adjusting the orbital period ($P$), the time of the periastron passage ($T_0$), the eccentricity ($e$), the angle of periastron ($\omega$), the systematic velocities ($\gamma$), and the semi-amplitudes ($K$). We averaged the radial velocity measurements of individual lines and approximated the error-bars of these values as the average error-bars coming from individual lines corrected with a factor of $\sqrt{n}$, where $n$ is the number of individual lines per spectra used in the radial velocity measurements. We have weighted the average radial velocities according to these errors as $w = 1/\sigma$ in the fit.

From the orbital parameters it is possible to derive -- as a function of the inclination angle -- the masses and the semi-major axes \citep[see, e.g.,][]{ramm_phd}. We give further notes and the results for the two systems in the subsections below.

\begin{table}
\caption{Orbital and physical parameters derived for the best-fit Keplerian model of KIC\,4931738, along with fundamental parameters and detailed abundances.}
\label{fundparams1}
\centering
\renewcommand{\arraystretch}{1.25}
\setlength{\tabcolsep}{1pt}
\begin{tabular}{l l@{\hskip 30pt} r c l c r c l}
\hline\hline
\multicolumn{2}{l}{Parameter} & \multicolumn{3}{c}{KIC\,4931738 A} && \multicolumn{3}{c}{KIC\,4931738 B} \\
\hline
&& \multicolumn{7}{c}{From fitting synthetic spectra}\\
\hline
\multicolumn{2}{l}{$P\,(\mathrm{days})$}&\multicolumn{3}{r}{$14.197$}&$\pm$&\multicolumn{3}{l}{$0.002$}\\
\multicolumn{2}{l}{$T_0\,(\mathrm{BJD})$}&\multicolumn{3}{r}{$2\,456\,059.904$}&$\pm$&\multicolumn{3}{l}{$0.015$}\\
\multicolumn{2}{l}{$e$}&\multicolumn{3}{r}{$0.191$}&$\pm$&\multicolumn{3}{l}{$0.001$}\\
\multicolumn{2}{l}{$\omega\,({\degr})$}&\multicolumn{3}{r}{$45.580$}&$\pm$&\multicolumn{3}{l}{$0.283$}\\
\multicolumn{2}{l}{$\gamma\,(\mathrm{km\,s}^{-1})$}&\multicolumn{3}{r}{$12.241$}&$\pm$&\multicolumn{3}{l}{$0.070$}\\
\multicolumn{2}{l}{$K\,(\mathrm{km\,s}^{-1})$}&$-62.717$&$\pm$&$0.476$&&$83.941$&$\pm$&$0.067$\\
\multicolumn{2}{l}{$\sigma_{O-C}\,(\mathrm{km\,s}^{-1})$}&$2.998$&&&&$0.301$&&\\
\multicolumn{2}{l}{$\mathcal{M}\sin^3 i_\mathrm{orb}\,(\mathcal{M}_{\sun})$}&$2.511$&$\pm$&$0.022$&&$1.876$&$\pm$&$0.029$\\
\multicolumn{2}{l}{$a\sin i_\mathrm{orb}\,(R_{\sun})$}&$17.275$&$\pm$&$0.134$&&$23.121$&$\pm$&$0.022$\\
\multicolumn{2}{l}{$q$}&\multicolumn{3}{r}{$0.747$}&$\pm$&\multicolumn{3}{l}{$0.018$}\\
\hline
&& \multicolumn{7}{c}{From spectral disentangling}\\
\hline
\multicolumn{2}{l}{$T_0\,(\mathrm{BJD})$}&\multicolumn{3}{r}{$2\,456\,059.910$}&$\pm$&\multicolumn{3}{l}{$0.029$}\\
\multicolumn{2}{l}{$e$}&\multicolumn{3}{r}{$0.190$}&$\pm$&\multicolumn{3}{l}{$0.004$}\\
\multicolumn{2}{l}{$\omega\,({\degr})$}&\multicolumn{3}{r}{$45.651$}&$\pm$&\multicolumn{3}{l}{$0.750$}\\
\multicolumn{2}{l}{$K\,(\mathrm{km\,s}^{-1})$}&$-62.547$&$\pm$&$0.431$&&$84.174$&$\pm$&$0.506$\\
\hline
&& \multicolumn{7}{c}{Initial estimate from restricted fitting} \\
\hline
\multicolumn{2}{l}{$T_\mathrm{eff}\,(\mathrm{K})$} & $14750$&$\pm$&$1000$ && $10750$&$\pm$&$1000$\\
\multicolumn{2}{l}{$\log g\,\mathrm{(cgs)}$} & $4.0$&$\pm$&$0.5$ && $4.5$&$\pm$&$0.5$ \\
\multicolumn{2}{l}{$\log Z/Z_{\sun}$} & $0.0$&$\pm$&$0.5$ && $0.0$&$\pm$&$0.5$ \\
\multicolumn{2}{l}{$v \sin i_\mathrm{rot}\,(\mathrm{km\,s}^{-1})$} & $9$&$\pm$&$2$ && $9$&$\pm$&$4$ \\
\hline
&& \multicolumn{7}{c}{Analysis of disentangled spectra} \\
\hline
\multicolumn{2}{l}{$T_\mathrm{eff}\,(\mathrm{K})$} & $13730$&$\pm$&$200$ && $11370$&$\pm$&$250$\\
\multicolumn{2}{l}{$\log g\,\mathrm{(cgs)}$} & $3.97$&$\pm$&$0.05$ && $4.37$&$\pm$&$0.10$ \\
\multicolumn{2}{l}{$\log Z/Z_{\sun}$} & $-0.24$&$\pm$&$0.10$ && $+0.10$&$\pm$&$0.10$ \\
\multicolumn{2}{l}{$v \sin i_\mathrm{rot}\,(\mathrm{km\,s}^{-1})$} & $10.5$&$\pm$&$1.0$ && $6.6$&$\pm$&$1.0$ \\
\multicolumn{2}{l}{$\xi_\mathrm{t}\,(\mathrm{km\,s}^{-1})$}& $2.0$&\multicolumn{2}{l}{(fixed)}&& $2.0$&\multicolumn{2}{l}{(fixed)} \\
\multicolumn{2}{l}{Spectral type\tablefootmark{a}}&\multicolumn{3}{c}{B6\,V}&&\multicolumn{3}{c}{B8.5\,V}\\
\hline
He&$[-1.11]$& $+0.10$&$\pm$&$0.10$ && $+0.00$&$\pm$&$0.10$ \\
Fe&$[-4.59]$& $-0.33$&$\pm$&$0.10$ && $+0.15$&$\pm$&$0.10$ \\
Mg&$[-4.51]$& $+0.00$&$\pm$&$0.12$ && $-0.08$&$\pm$&$0.20$ \\
Si&$[-4.53]$& $-0.30$&$\pm$&$0.15$ && $+0.15$&$\pm$&$0.20$ \\
Ti&$[-7.14]$& $-0.40$&$\pm$&$0.35$ && $-0.10$&$\pm$&$0.20$ \\
Cr&$[-6.40]$& $-0.30$&$\pm$&$0.35$ && $+0.00$&$\pm$&$0.25$ \\
\hline
\end{tabular}
\tablefoot{All abundances are given in units of $\log(\mathrm{N/N_{tot}})$, relative to the Solar values listed by \citet{2005ASPC..336...25A} which are also indicated in parentheses after each element. The mass ratio is defined as $q=\mathcal{M}_\mathrm{B}/\mathcal{M}_\mathrm{A}$. \tablefoottext{a}{Spectral type have been determined based on $T_\mathrm{eff}$ and $\log g$ values by using an interpolation in the tables given by \citet{1982SchmidtKalerBook}.}}
\end{table}

\begin{table}
\caption{Orbital and physical parameters derived for the best-fit Keplerian model of KIC\,6352430 (using observations from 2012), along with fundamental parameters and detailed abundances.}
\label{fundparams2}
\centering
\renewcommand{\arraystretch}{1.25}
\setlength{\tabcolsep}{1pt}
\begin{tabular}{l l@{\hskip 30pt} r c l c r c l}
\hline\hline
\multicolumn{2}{l}{Parameter} & \multicolumn{3}{c}{KIC\,6352430 A} && \multicolumn{3}{c}{KIC\,6352430 B} \\
\hline
&& \multicolumn{7}{c}{From fitting synthetic spectra}\\
\hline
\multicolumn{2}{l}{$P\,(\mathrm{days})$}&\multicolumn{3}{r}{$26.551$}&$\pm$&\multicolumn{3}{l}{$0.019$}\\
\multicolumn{2}{l}{$T_0\,(\mathrm{BJD})$}&\multicolumn{3}{r}{$2\,455\,486.518$}&$\pm$&\multicolumn{3}{l}{$0.501$}\\
\multicolumn{2}{l}{$e$}&\multicolumn{3}{r}{$0.370$}&$\pm$&\multicolumn{3}{l}{$0.003$}\\
\multicolumn{2}{l}{$\omega\,({\degr})$}&\multicolumn{3}{r}{$148.755$}&$\pm$&\multicolumn{3}{l}{$0.587$}\\
\multicolumn{2}{l}{$\gamma\,(\mathrm{km\,s}^{-1})$}&$-24.838$&$\pm$&$0.342$&&$-24.530$&$\pm$&$0.211$\\
\multicolumn{2}{l}{$K\,(\mathrm{km\,s}^{-1})$}&$-36.595$&$\pm$&$0.571$&&$83.715$&$\pm$&$0.344$\\
\multicolumn{2}{l}{$\sigma_{O-C}\,(\mathrm{km\,s}^{-1})$}&$1.976$&&&&$0.412$&&\\
\multicolumn{2}{l}{$\mathcal{M}\sin^3 i_\mathrm{orb}\,(\mathcal{M}_{\sun})$}&$2.673$&$\pm$&$0.063$&&$1.168$&$\pm$&$0.041$\\
\multicolumn{2}{l}{$a\sin i_\mathrm{orb}\,(R_{\sun})$}&$17.842$&$\pm$&$0.299$&&$40.816$&$\pm$&$0.215$\\
\multicolumn{2}{l}{$q$}&\multicolumn{3}{r}{$0.437$}&$\pm$&\multicolumn{3}{l}{$0.027$}\\
\hline
&& \multicolumn{7}{c}{From spectral disentangling}\\
\hline
\multicolumn{2}{l}{$T_0\,(\mathrm{BJD})$}&\multicolumn{3}{r}{$2\,455\,486.563$}&$\pm$&\multicolumn{3}{l}{$0.021$}\\
\multicolumn{2}{l}{$e$}&\multicolumn{3}{r}{$0.371$}&$\pm$&\multicolumn{3}{l}{$0.003$}\\
\multicolumn{2}{l}{$\omega\,({\degr})$}&\multicolumn{3}{r}{$149.425$}&$\pm$&\multicolumn{3}{l}{$0.359$}\\
\multicolumn{2}{l}{$K\,(\mathrm{km\,s}^{-1})$}&$-36.411$&$\pm$&$1.569$&&$83.854$&$\pm$&$0.403$\\
\hline
&& \multicolumn{7}{c}{Initial estimate from restricted fitting} \\
\hline
\multicolumn{2}{l}{$T_\mathrm{eff}\,(\mathrm{K})$} & $13000$&$\pm$&$750$ && $7500$&$\pm$&$1000$\\
\multicolumn{2}{l}{$\log g\,\mathrm{(cgs)}$} & $4.0$&$\pm$&$0.5$ && $4.5$&$\pm$&$0.5$ \\
\multicolumn{2}{l}{$\log Z/Z_{\sun}$} & $0.0$&$\pm$&$0.5$ && $-0.5$&$\pm$&$0.25$ \\
\multicolumn{2}{l}{$v \sin i_\mathrm{rot}\,(\mathrm{km\,s}^{-1})$} & $70$&$\pm$&$5$ && $12$&$\pm$&$4$ \\
\hline
&& \multicolumn{7}{c}{Analysis of disentangled spectra} \\
\hline
\multicolumn{2}{l}{$T_\mathrm{eff}\,(\mathrm{K})$} & $12810$&$\pm$&$200$ && $6805$&$\pm$&$100$\\
\multicolumn{2}{l}{$\log g\,\mathrm{(cgs)}$} & $4.05$&$\pm$&$0.05$ && $4.26$&$\pm$&$0.15$ \\
\multicolumn{2}{l}{$\log Z/Z_{\sun}$} & $-0.13$&$\pm$&$0.07$ && $-0.33$&$\pm$&$0.10$ \\
\multicolumn{2}{l}{$v \sin i_\mathrm{rot}\,(\mathrm{km\,s}^{-1})$} & $69.8$&$\pm$&$2.0$ && $9.8$&$\pm$&$1.0$ \\
\multicolumn{2}{l}{$\xi_\mathrm{t}\,(\mathrm{km\,s}^{-1})$}& $2.0$&\multicolumn{2}{l}{(fixed)}&& $2.0$&\multicolumn{2}{l}{(fixed)} \\
\multicolumn{2}{l}{Spectral type\tablefootmark{a}}&\multicolumn{3}{c}{B7\,V}&&\multicolumn{3}{c}{F2.5\,V}\\
\hline
He&$[-1.11]$& $+0.06$&$\pm$&$0.10$ &&&& \\
Fe&$[-4.59]$& $-0.25$&$\pm$&$0.10$ && $-0.30$&$\pm$&$0.10$ \\
Mg&$[-4.51]$& $+0.14$&$\pm$&$0.10$ && $-0.35$&$\pm$&$0.20$ \\
Si&$[-4.53]$& $-0.20$&$\pm$&$0.15$ && $-0.35$&$\pm$&$0.30$ \\
Ti&$[-7.14]$& $-0.40$&$\pm$&$0.35$ && $-0.25$&$\pm$&$0.15$ \\
Cr&$[-6.40]$& $-0.20$&$\pm$&$0.35$ && $-0.10$&$\pm$&$0.20$ \\
Ca&$[-5.73]$&&&&&                     $-0.45$&$\pm$&$0.25$ \\
Sc&$[-8.99]$&&&&&                     $-0.10$&$\pm$&$0.35$ \\
Mn&$[-6.65]$&&&&&                     $-0.30$&$\pm$&$0.35$ \\
Y& $[-9.83]$&&&&&                     $-0.20$&$\pm$&$0.35$ \\
Ni&$[-5.81]$&&&&&                     $-0.50$&$\pm$&$0.25$ \\
\hline
\end{tabular}
\tablefoot{We refer to Table\,\ref{fundparams1} for notes and explanation.}
\end{table}

\begin{figure}
\resizebox{\hsize}{!}{\includegraphics{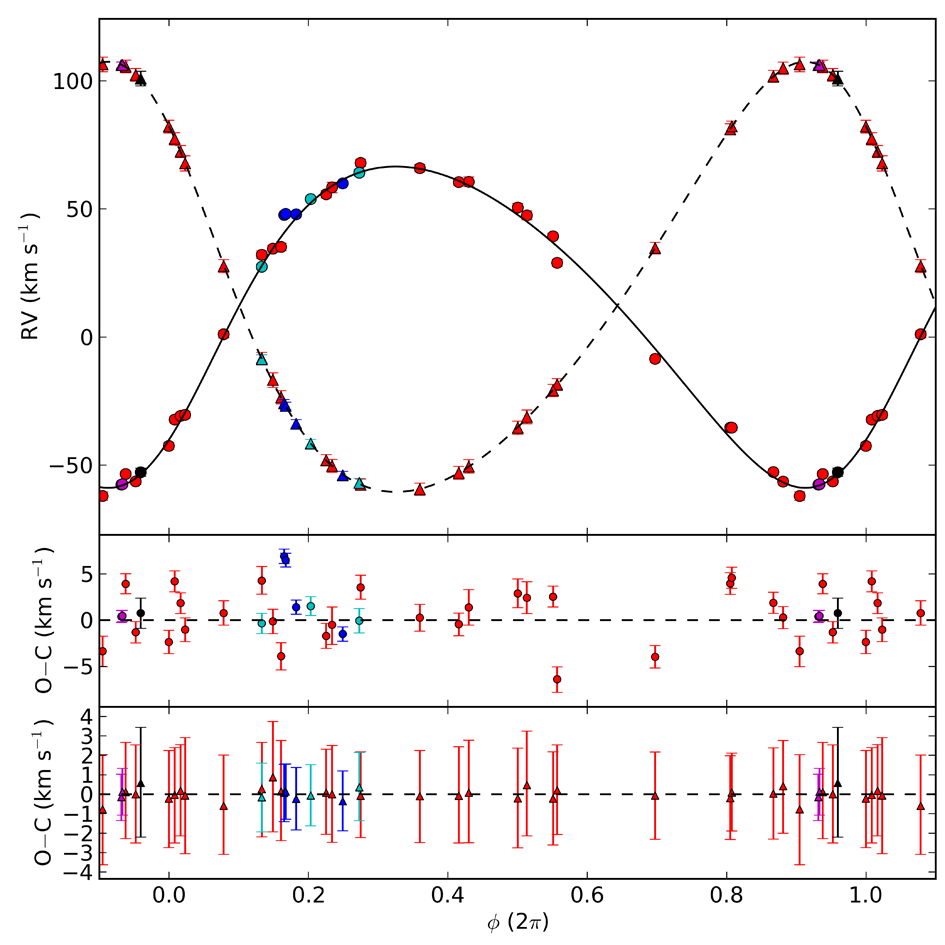}} 
\caption{(\textit{upper panel}) Spectroscopic orbital solution for KIC\,4931738 (solid line: primary, dashed line: secondary) and the measured radial velocities. Filled circles indicate measurements of the primary, filled triangles denote measurements of the secondary component using \textsc{Hermes} (red markers), \textsc{Fies} (blue markers), \textsc{Hamilton} (cyan markers), \textsc{Arces} (magenta markers), and \textsc{Isis} (black markers). (\textit{middle panel}) Residuals of the primary component. (\textit{lower panel}) Residuals of the secondary component.}
\label{orbit1}
\end{figure}

\begin{figure}
\resizebox{\hsize}{!}{\includegraphics{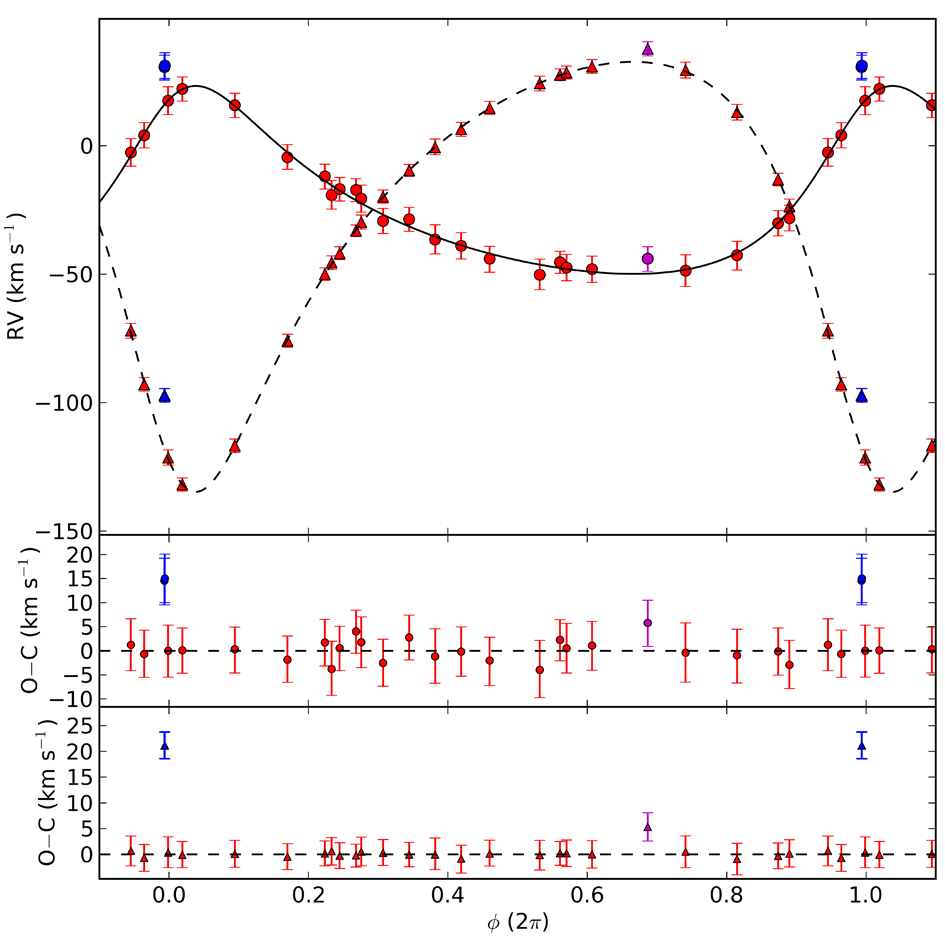}} 
\caption{(\textit{upper panel}) Spectroscopic orbital solution for KIC\,6352430 (solid line: primary, dashed line: secondary) and the measured radial velocities. Filled circles indicate measurements of the primary, filled triangles denote measurements of the secondary component. Observations from 2012 are plotted with red symbols, while measurements from 2011 and 2010 are plotted with magenta and blue markers, respectively. (\textit{middle panel}) Residuals of the primary component. (\textit{lower panel}) Residuals of the secondary component.}
\label{orbit2}
\end{figure}

\subsubsection{KIC\,4931738}

During the fit we left out four measurements (one \textsc{Hermes} and three \textsc{Fies} exposures) where the components had nearly the same radial velocities such as the components were mixed up during the automated radial velocity determination due to the overlapping lines, as well as one measurement coming from the blue arm of \textsc{Isis}, where it was not possible to determine the zero point offset. The first fit showed that there is an order of magnitude higher scatter in the radial velocity measurements of the primary. We interpret this as a result of the underlying pulsations. For this reason we have redone the fit by first fitting the orbit of the secondary component, then fixing these values and fitting the remaining $K_1$ for the final parameters. This way we determined a binary orbit with a period of $14.197\pm0.002$ days and an eccentricity of $0.191\pm0.001$. The orbital and physical parameters derived from spectroscopy are listed in Table\,\ref{fundparams1}, while the radial velocity curves and the best fit solutions are plotted in Fig.\,\ref{orbit1}.

\begin{figure}
\resizebox{\hsize}{!}{\includegraphics{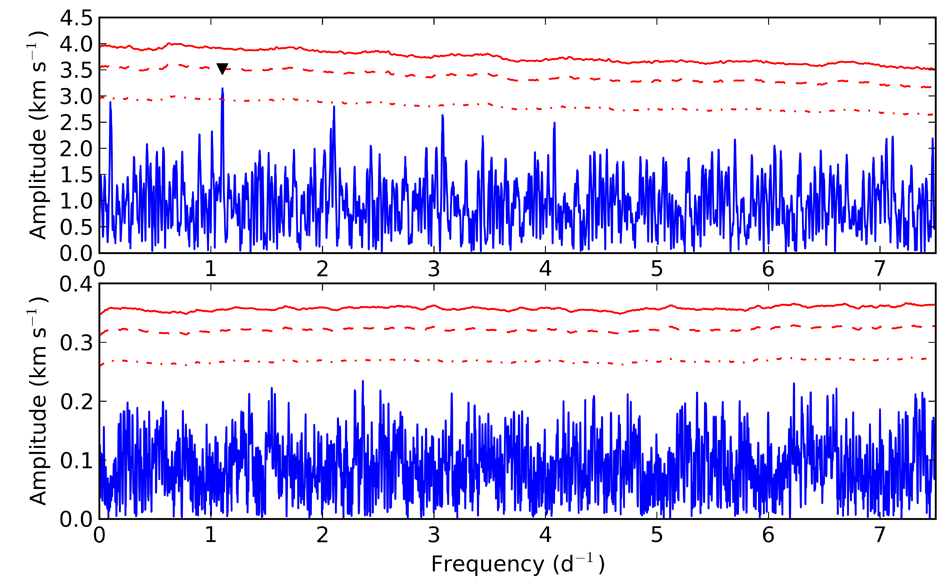}} 
\caption{The Scargle periodogram of the $O-C$ values of the primary (\textit{upper panel}) and secondary component (\textit{lower panel}) of KIC\,4931738. The significance levels corresponding to a signal-to-noise ratio (calculated in a $3\,\mathrm{d}^{-1}$ window) of 4, 3.6, and 3 are plotted with red solid, dashed, and dot-dashed lines, respectively. The peak at $1.1040\pm0.0006\,\mathrm{d}^{-1}$ is marked with a black triangle.}
\label{rvresidualfourier1}
\end{figure}

\begin{figure}
\resizebox{\hsize}{!}{\includegraphics{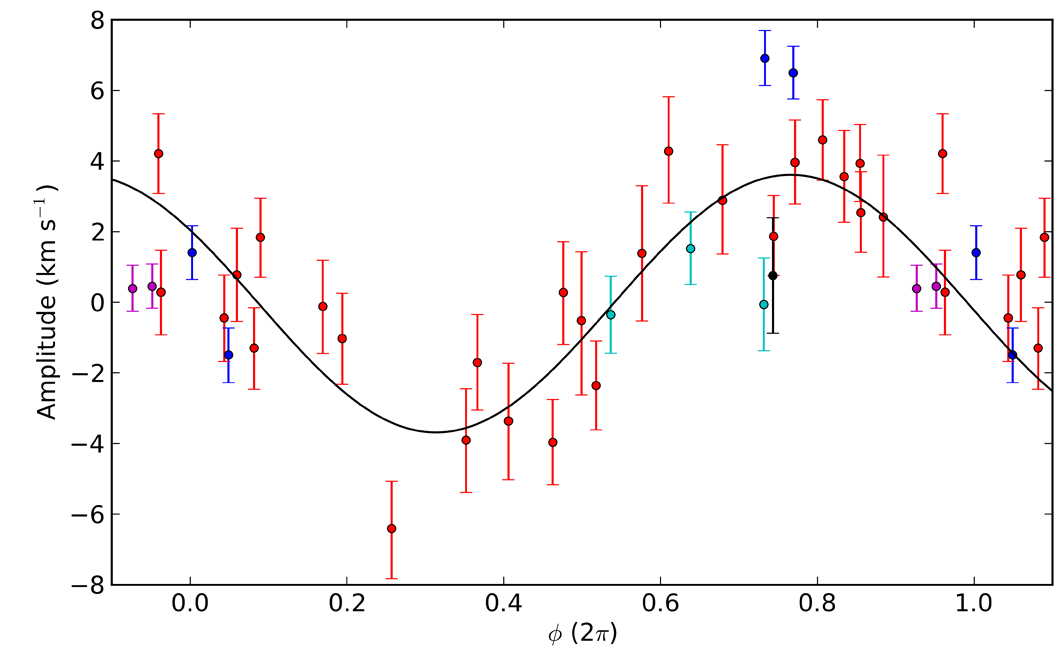}} 
\caption{$O-C$ values of the primary component of KIC\,4931738 phased with $f_2 = 1.105147\,\mathrm{d}^{-1}$. A sinusoidal fit is shown with a solid black line. For the explanation of different colours we refer to Fig.\,\ref{orbit1}.}
\label{rvresidualphaseplot}
\end{figure}

The Scargle periodogram \citep{1982ApJ...263..835S} of the residuals of the radial velocities of the primary ($O_1-C_1$) shows one clear peak (and its $1\,\mathrm{d}^{-1}$ aliases -- see Fig.\,\ref{rvresidualfourier1}) above the noise level at $1.1040\pm0.0006\,\mathrm{d}^{-1}$ with an amplitude of $3.4\pm0.5\,\mathrm{km\,s}^{-1}$, and a signal-to-noise value of 3.2 (calculated in a $3\,\mathrm{d}^{-1}$ window centred on the peak). Although this is below the generally accepted significance level of 4, we will show in Sect.\,\ref{frequencynotes1} that this is one of the strongest frequencies ($f_2$ in Table\,\ref{combinations1}) found in the \textit{Kepler} photometry of the star. The phase diagram of the $O_1-C_1$ values calculated using $f_2$ (see Fig.\,\ref{rvresidualphaseplot}) supports the suggested pulsational origin of the observed scatter, and also connects the pulsations seen in the light curve to the primary star. We also checked the $O_1-C_1$ diagram using $f_1$ (in Table\,\ref{combinations1}) as well and found nothing. We note that heat driven oscillation modes can have very different amplitudes in photometry, which is determined by the temperature variation, and in spectroscopy, which diagnoses the velocity variation. This can be the reason why two modes of the same $l$ and $m$ with similar photometric amplitudes (such as $f_1$ and $f_2$ -- see Sect.\,\ref{seismicint1} for more information on these modes) appear with different radial velocity amplitudes. There is no sign of periodic variability in the residuals of the secondary.

\subsubsection{KIC\,6352430}

Thanks to the difference in the spectral type of the components and their projected rotational velocities, there was no need to drop measurements where the two components have nearly the same radial velocity. Our method works well in these circumstances, because there is no risk of misidentification of components during the fitting of synthetic spectra, as there was in the case of KIC\,4931738. The first fits using the full data set brought our attention to the slight discrepancy between measurements made in 2012 and the three measurements in the previous two seasons. To resolve this issue we made the final fit using only the data points from 2012, which delivered a very well constrained orbit with an orbital period of $26.551\pm0.019$ days and an eccentricity of $0.370\pm0.003$. The orbital and physical parameters derived from spectroscopy are listed in Table\,\ref{fundparams2}, while the radial velocity curves and the best fit solutions are plotted in Fig.\,\ref{orbit2}.

As it can be seen in Fig.\,\ref{orbit2}, measurements from 2010 and 2011 are systematically and significantly off this fit, which we explain by the presence of a third body in a much wider orbit around the centre of mass. To put constraints on that component, further observations would be necessary. There is no significant frequency in the Scargle periodogram of the residuals.

\subsection{Spectral disentangling}
As introduced by \citet{1994A&A...281..286S}, in the method of spectral disentangling, one simultaneously solves for the individual spectra of stellar components of a multiple system and set of the orbital elements. \citet{1995A&AS..114..393H} suggested an application of the technique in Fourier space which is significantly faster compared to the disentangling in wavelength domain. In this work, we apply the spectral disentangling technique in Fourier space as implemented in the FDBinary code \citep{2004ASPC..318..111I}.

We have used four spectral intervals centred on sufficiently strong helium and metal lines (\ion{He}{i} at $4471\AA$ and \ion{Mg}{ii} at $4481\AA$; \ion{He}{i} at $4921\AA$; \ion{He}{i} at $5016\AA$ and \ion{Si}{ii} at $5041\AA$; \ion{Mg}{i} at $5167\AA$ and \ion{Fe}{ii} at $5169\AA$) for measuring the orbital elements. The final mean orbital elements and $1$-$\sigma$ error bars are in perfect agreement with the values determined from fitting synthetic composite spectra in Sect.\,\ref{rvmeasurement}, and are listed in the bottom section of Table\,\ref{fundparams1} and Table\,\ref{fundparams2}. Because the uncertainty in the normalization of the observed spectra raises towards the blue edge of the spectra, we restricted our spectral decomposition to the wavelength range with $\lambda > 4050\AA$. It is well-known that spectral disentangling in Fourier space suffers from undulations in the resulting decomposed spectra \citep[see, e.g.,][]{1995A&AS..114..393H,2004ASPC..318..111I}. To overcome this problem, we divided our spectra into $\sim50\AA$ overlapping wavelength regions and corrected continua of the resulting decomposed spectra in these short regions using spline functions. An exception has been made for wide Balmer lines for which the decomposition was done for the entire profile. The final, corrected decomposed spectra for each component have been obtained by merging all the short and Balmer lines disentangled intervals. 

\subsection{Fundamental parameters and abundances}\label{fundparsandabundances}
For the analysis of the decomposed spectra, we used the GSSP package \citep{2012MNRAS.422.2960T,2011A&A...526A.124L}. The code relies on a comparison between observed and synthetic spectra computed in a grid of $T_\mathrm{eff}$, $\log g$, $\xi_\mathrm{t}$, $[M/H]$, and $v \sin i$, and finds the optimum values of these parameters from a minimum in $\chi^2$. Besides that, individual abundances of different chemical species can be adjusted in the second step assuming a stellar atmosphere model of a certain global metallicity. The error bars are represented by $1$-$\sigma$ confidence levels computed from $\chi^2$ statistics. The grid of atmosphere models has been computed using the most recent version of the LLmodels code \citep{2004A&A...428..993S}. For the calculation of synthetic spectra, we use the LTE-based code SynthV \citep{1996ASPC..108..198T} which allows to compute the spectra based on individual elemental abundances.

\begin{figure*}
\resizebox{\hsize}{!}{\includegraphics{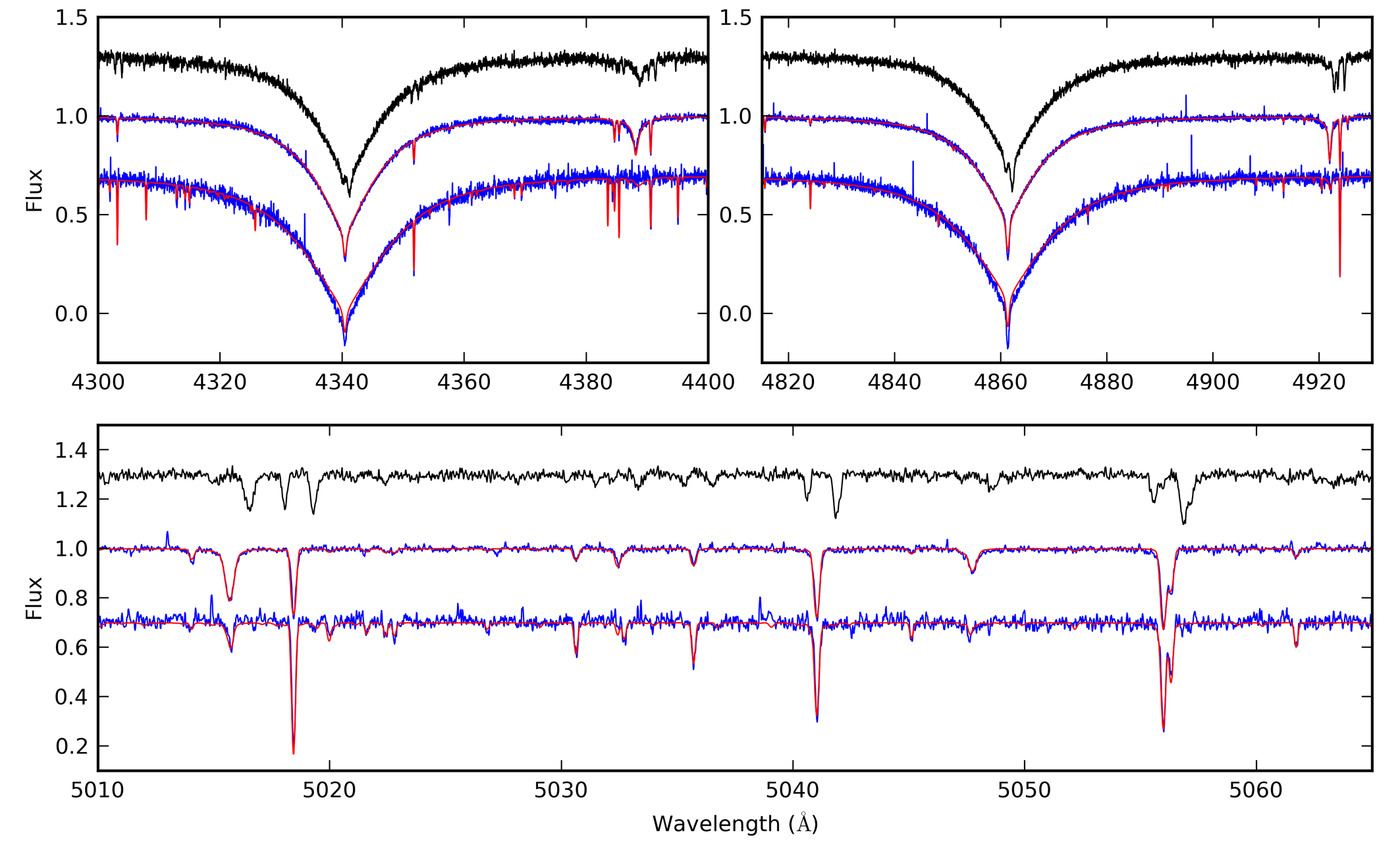}} 
\caption{Comparison of normalised observed, disentangled and synthetic spectra in different wavelength regions for KIC\,4931738. In each panel, the highest signal-to-noise \textsc{Fies} spectrum is plotted with a black solid line shifted upwards by 0.3 flux units, the observed and synthetic spectra of the primary component are plotted with blue and red solid lines, respectively, and the observed and synthetic spectra of the secondary component are plotted in a similar manner, but shifted downwards by 0.3 flux units.}
\label{spectralfit1}
\end{figure*}

\begin{figure*}
\resizebox{\hsize}{!}{\includegraphics{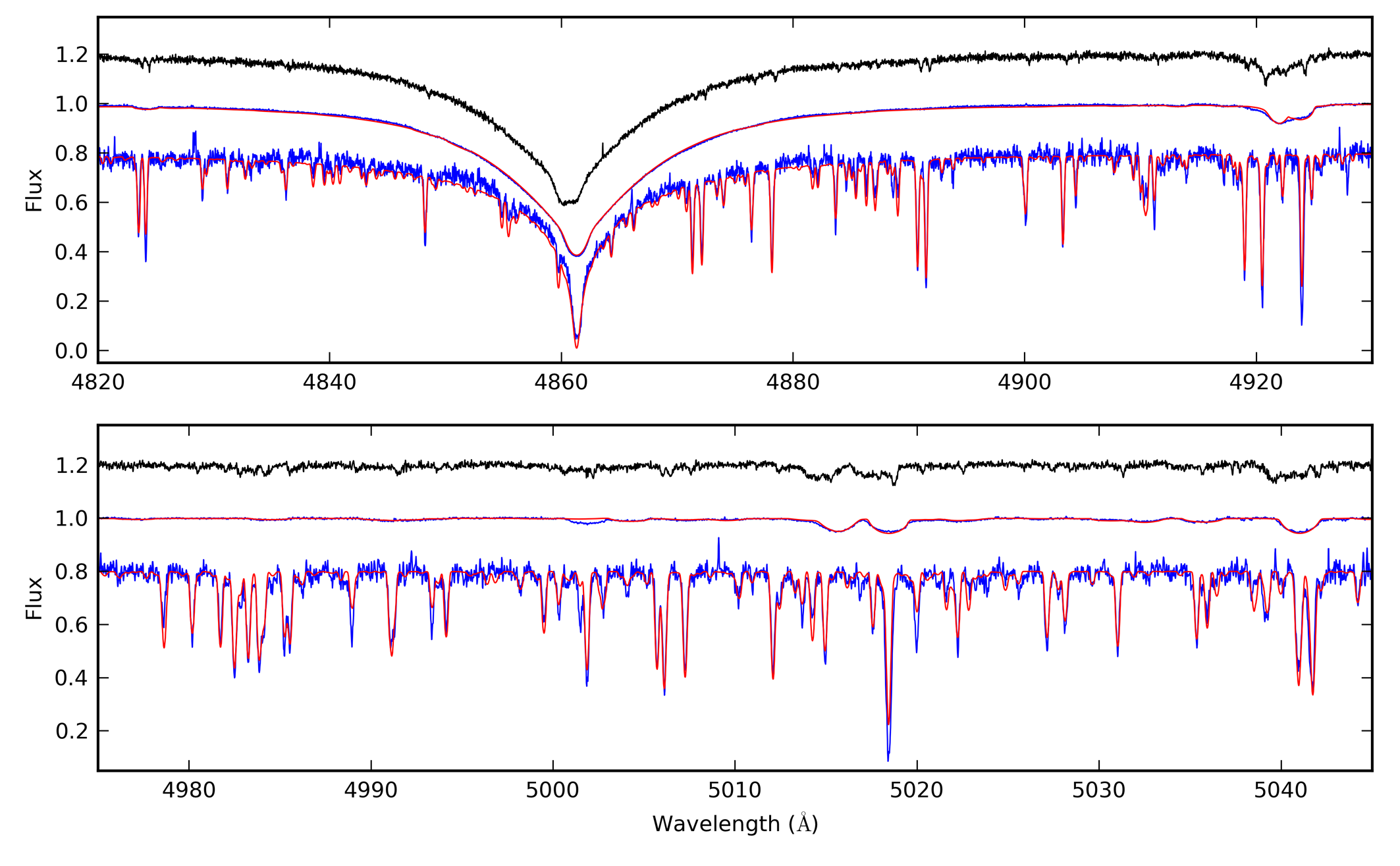}} 
\caption{Comparison of normalised observed, disentangled and synthetic spectra in different wavelength regions for KIC\,6352430. In each panel, the highest signal-to-noise \textsc{Hermes} spectrum is plotted with a black solid line shifted upwards by 0.2 flux units, the observed and synthetic spectra of the primary component are plotted with blue and red solid lines, respectively, and the observed and synthetic spectra of the secondary component are plotted in a similar manner, but shifted downwards by 0.2 flux units.}
\label{spectralfit2}
\end{figure*}

The bottom sections of Table\,\ref{fundparams1} and Table\,\ref{fundparams2} list the atmospheric parameters and elemental abundances for the individual stellar components of KIC\,4931738 and KIC\,6352430. The spectral types and the luminosity classes have been derived by interpolating in the tables published by \citet{1982SchmidtKalerBook}. Fig.\,\ref{spectralfit1} and Fig.\,\ref{spectralfit2} show the quality of the observed and disentangled spectra and their fit for both components of both binary systems in different wavelength regions.

%%%%%%%%%%%%%%%%%%
%%%Kepler photo%%%
%%%%%%%%%%%%%%%%%%

\section{\textit{Kepler} photometry}\label{photometry}
\subsection{Data reduction}\label{data}
Due to the quarterly rolls of the \textit{Kepler} satellite which are performed to keep the solar panels facing the Sun along the course of the $\sim370$ day orbit, the photometric data is delivered in quarters (Q). Most data are obtained in long-cadence (LC) mode with an integration time of  29.4 minutes. The characteristics of the LC mode are described by \citet{2010ApJ...713L.120J}. In this paper we use long-cadence data from Q0 to Q13 (March 2009 to June 2012), which gives a timebase of slightly over three calendar years, which is nearly an order of magnitude longer than the one typically provided by CoRoT observations. Q6 and Q10 are missing for KIC\,4931738 due to the malfunctioning of Module 3 of the camera, while Q0 is not available for KIC\,6352430.

\begin{figure}
\resizebox{\hsize}{!}{\includegraphics{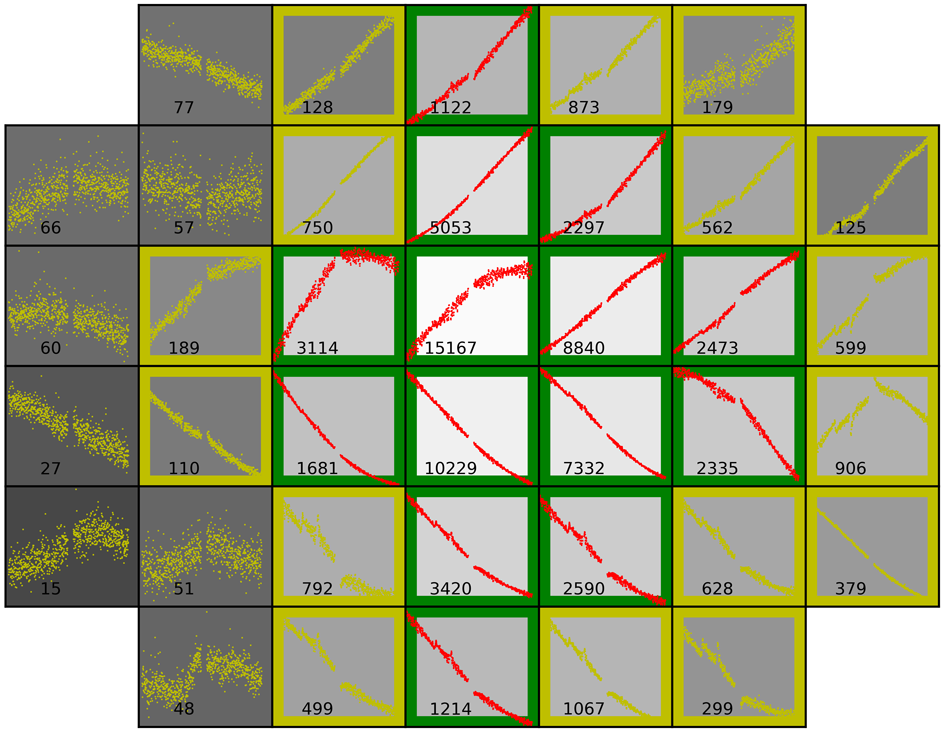}} 
\caption{Pixel mask for the Q8 data of KIC\,4931738. The light curve is plotted for every individual pixel that was downloaded from the spacecraft. The value in every pixel as well as the background colour, indicate the signal to noise ratio of the flux in the pixel. The pixels with green borders were used to extract the standard \textit{Kepler} light curves. We have added the yellow pixels (significant signal is present) in our custom mask for the light curve extraction. This results in a light curve with significantly less instrumental effects than the standard extraction.}
\label{mask1}
\end{figure}

\begin{figure}
\resizebox{\hsize}{!}{\includegraphics{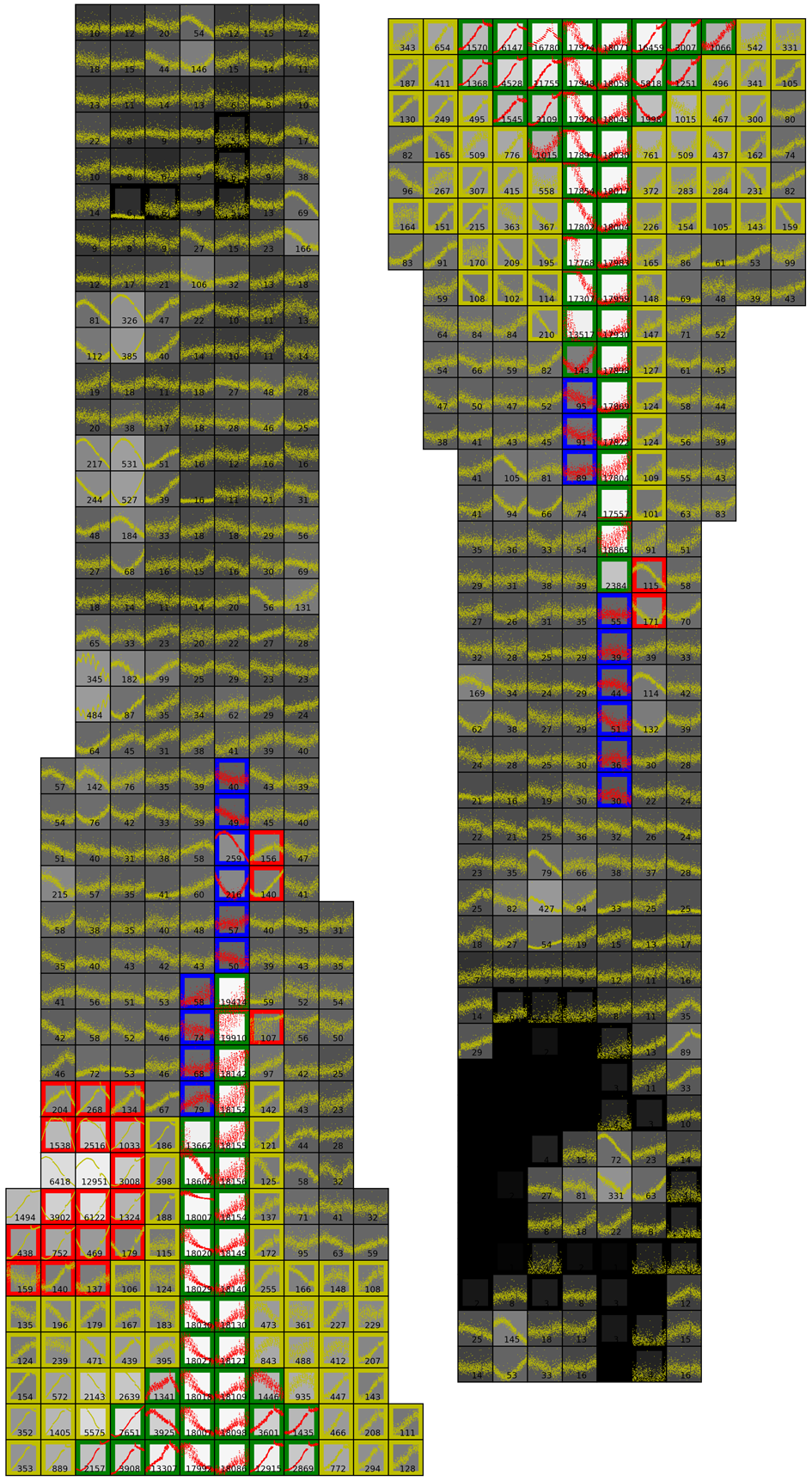}} 
%\resizebox{10 cm}{!}{\includegraphics{figure_kic6352430_mask.png}} 
\caption{Pixel mask for the Q3 data of KIC\,6352430, plotted in a similar manner to Fig.\,\ref{mask1}, and cut in two parts for better visibility. We have not used the blue pixels (not enough flux, typically around bright stars where the overflow from saturated pixels was less than expected) but have added the yellow pixels (significant signal is present) in our custom mask for the light curve extraction. Red pixels were flagged as being contaminated by flux from a background star and were not used despite the high flux level, while black pixels have too many NaN values.}
\label{mask2}
\end{figure}

Instead of using the standard pre-extracted light curves delivered through MAST, we have extracted the light curves from the pixel data information using custom masks. Our masks are determined for each quarter separately, and contain all pixels with significant flux. Compared to the standard masks, which contain only the pixels with a very high signal to noise ratio, including lower signal-to-noise pixels results in light curves with significantly less instrumental trends than the standard light curves. The automated pixel selection process and the benefits in terms of reducing instrumental effects are explained in Bloemen et al. (in preparation). Examples of the sets of pixels we used are shown in Fig.\,\ref{mask1} and Fig.\,\ref{mask2}.

We have manually removed obvious outliers from the data. To correct for the different flux levels of the individual quarters (which is a result of the targets being observed by different CCDs -- modules -- of the CCD array after each quarterly roll, having slightly different response curves and pixel masks) we have divided them by a fitted second order polynomial. As after these initial steps there were no clear jumps or trends visible anymore, we merged the quarters and used the resulting datasets in our analysis.

\subsection{Frequency analysis}\label{frequanal}
To derive the frequencies present in the light variation of the targets we performed an iterative prewhitening procedure whose description is already given by \citet{2009A&A...506..111D} and thus omitted here. This provided a list of amplitudes ($A_j$), frequencies ($f_j$), and phases ($\theta_j$), by which the light curve can be modelled via $n_f$ frequencies in the well-known form of \[F(t_i)=c+\sum_{j=1}^{n_f}A_j\sin[2\pi(f_j t_i + \theta_j)].\] The prewhitening procedure was stopped when a $p$ value of $p = 0.001$ was reached in hypothesis testing.

As shown by \citet{2012A&A...542A..55P} the way one defines the significance criteria can strongly affect the final number of accepted frequencies in the Fourier-analysis. Here we decided to use the classical approach \citep{1993A&A...271..482B} and measure the signal-to-noise value (SNR) as the ratio of a given peak to the average residual amplitude in a $3\,\mathrm{d}^{-1}$ window around the peak, and accept a peak as significant when $\mathrm{SNR}\ge4$. We used this set of frequencies to model the light curve. 

From the numbers listed in Sect.\,\ref{frequencynotes1} and Sect.\,\ref{frequencynotes2} it is clear that using this criterion, in the lower frequency ranges, the periodogram is basically saturated in frequency, which means that the average separation of significant peaks is close, or even less than the theoretical frequency resolution (within this, two close peaks will appear in the periodogram with a frequency which is affected by the other peak and vice versa), which is given by the \citet{1978Ap&SS..56..285L} criterion of $\sim2.5/T = 0.0022\,\mathrm{d}^{-1}$. Even though these peaks meet the classical signal-to-noise criterion, and they are needed to properly model the light curve, we must restrict ourselves to a more significant subset of these frequencies, to avoid confusion in our tests later on. For this purpose, we define a selection criterion which requires a peak to have $\mathrm{SNR}\ge4$ in a $1\,\mathrm{d}^{-1}$ window to be selected. We will use these \textit{selected peaks} in our tests later on.

In the following paragraphs we describe the analysis of the frequencies found in the \textit{Kepler} photometry for the two binary systems.

\subsection{KIC\,4931738}\label{frequencynotes1}
\begin{figure*}
\resizebox{\hsize}{!}{\includegraphics{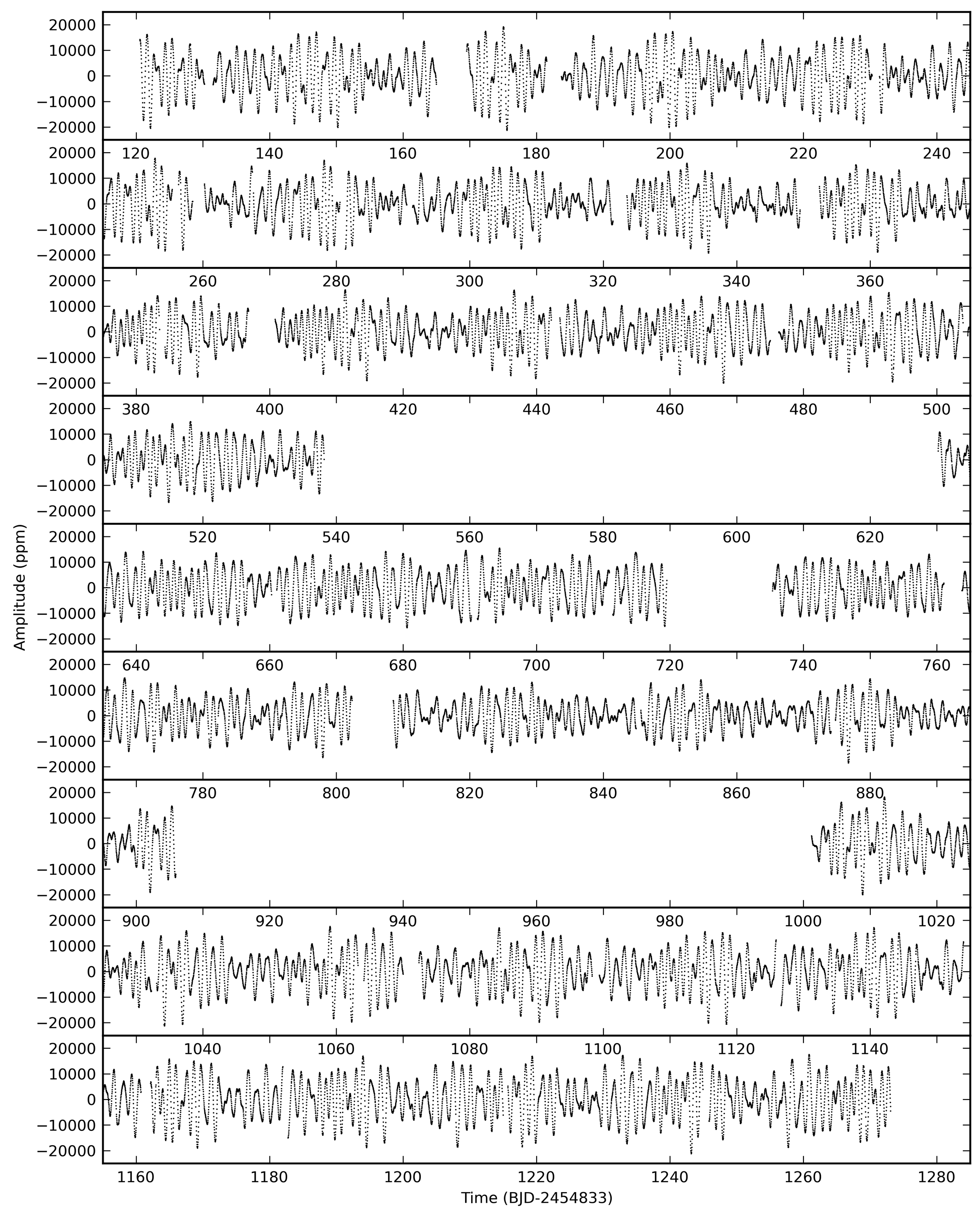}} 
\caption{The full reduced \textit{Kepler} light curve (black dots, brightest at the top) of KIC\,4931738. The absence of Q6 and Q10 data is clearly visible in the regions centred around 580 and 960 days.}
\label{lightcurve1}
\end{figure*}

The cleaned and detrended \textit{Kepler} light curve (see Fig.\,\ref{lightcurve1}) containing 43578 data points covers 1152.52 days starting from BJD 2454953.53821024, and has a duty cycle of 77.26\% due to the two quarter-long gaps caused by the malfunction (and loss) of Module 3 during Q4.

\begin{figure*}
\resizebox{\hsize}{!}{\includegraphics{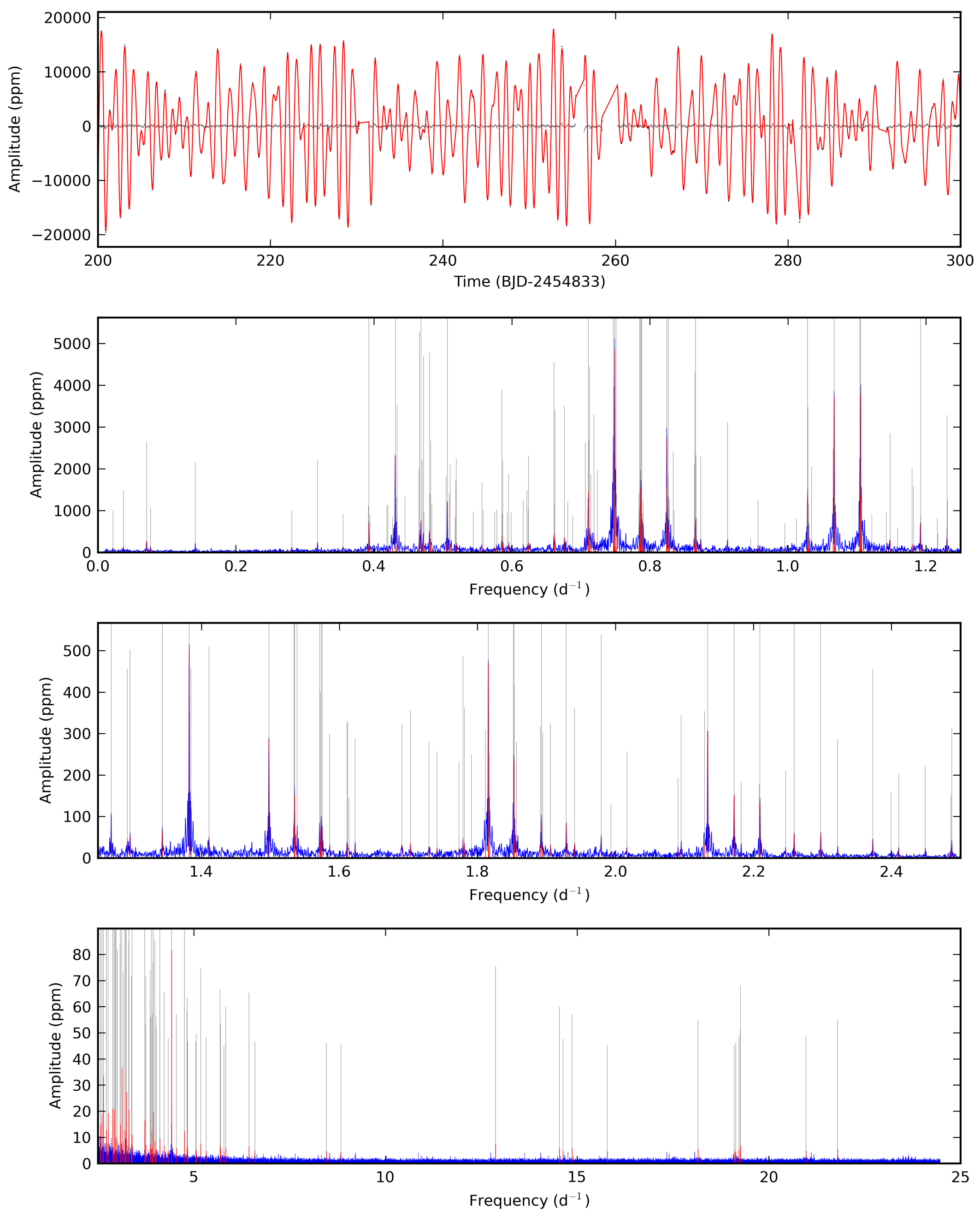}} 
\caption{(\textit{upper panel}) A zoom-in section of the reduced \textit{Kepler} light curve (black dots, brightest at the top) and residuals (gray dots) of KIC\,4931738 after prewhitening with a model (red solid line) constructed using the set of 1082 significant frequencies -- see text for further explanation. (\textit{lower three panels}) The Scargle periodogram of the full \textit{Kepler} light curve (blue solid line) showing the 262 selected frequencies (red vertical lines). For better visibility, the red lines are repeated in gray in the background, after multiplying their amplitude with a factor ten, and the signal from outside the plotted ranges is prewhitened for each panel.}
\label{fourier1}
\end{figure*}

The iterative prewhitening process returned 1784 frequencies of which 1082 met our significance criteria. The Scargle periodogram is shown in Fig.\,\ref{fourier1}. The model constructed using this set of frequencies provides a variance reduction of 99.94\% while bringing down the average signal levels from 213.4--135.3--4.5--3.8--3.7 ppm to 2.0--1.9--0.6--0.3--0.4 ppm, measured in $2\,\mathrm{d}^{-1}$ windows centred around 1, 2, 5, 10, and $20\,\mathrm{d}^{-1}$, respectively. Most of the power is distributed between 0.25 and $2.5\,\mathrm{d}^{-1}$: 887 peaks contribute to a total power density (integral power of significant peaks normalised with interval width) of $4.07\times10^7\,\mathrm{ppm}^2\mathrm{d}$, while there are only 40 peaks between 2.5 and $5\,\mathrm{d}^{-1}$ ($6.22\times10^3\,\mathrm{ppm}^2\mathrm{d}$), and 25 peaks above $5\,\mathrm{d}^{-1}$ ($38.6\,\mathrm{ppm}^2\mathrm{d}$). The 130 peaks below $0.25\,\mathrm{d}^{-1}$ provide a power density of $1.00\times10^6\,\mathrm{ppm}^2\mathrm{d}$. 262 of these frequencies met the selection criterion and are listed in Table\,\ref{frequtable1}.

The low frequency signal (especially the lowest frequencies) partly comes from minor and unavoidable instrumental trends left in the data (not 100\% perfect merging of the quarters, long term CCD stability issues, guiding issues, etc.), but the other part still has a physical origin, most probably granulation noise and/or low amplitude pulsations.

The binary orbital period of $14.197\pm0.002$ days ($f_\mathrm{SB2} = 0.0704374\pm0.0000099\,\mathrm{d}^{-1}$) can also be found in the frequency analysis of the light curve at $0.070457\pm0.000014\,\mathrm{d}^{-1}$ with an amplitude of $264\pm8\,\mathrm{ppm}$, along with twice the binary orbital frequency at $0.140869\pm0.000016\,\mathrm{d}^{-1}$ ($216\pm7\,\mathrm{ppm}$), which are the two strongest peaks below $0.25\,\mathrm{d}^{-1}$. The appearance of the harmonics is a sign of a slight ellipsoidal modulation. There is no sign of eclipses in the light curve phased with the orbital period. The light variation due to oscillations is too complicated to be prewhitened and interpret the photometric variability due to the orbital motion alone.

\subsubsection{Spacings in Fourier space}\label{spacings1}

\begin{figure}
\resizebox{\hsize}{!}{\includegraphics{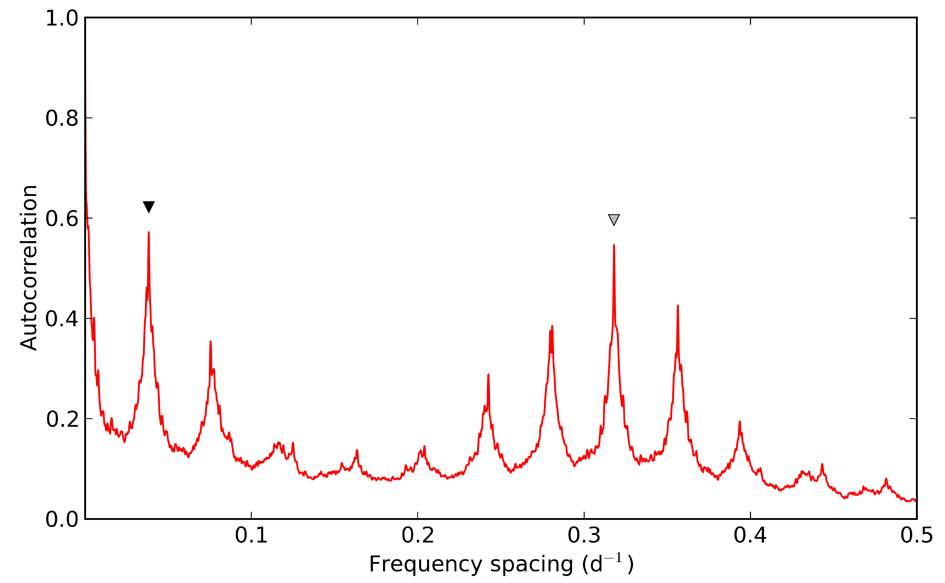}} 
\caption{Autocorrelation function (red solid line) of the Scargle periodogram of KIC\,4931738 between 0 and $5\,\mathrm{d}^{-1}$. The two strongest peaks are marked with black and grey triangles.}
\label{autocorroffourier1}
\end{figure}

In general, the Fourier spectrum is remarkably rich and structured. There is a clear comb-like structure between 0.3 and $2.3\,\mathrm{d}^{-1}$, containing alternating regions of high and low amplitude peak groups. Both these regions and the peak groups within them are equidistantly spaced. This can be clearly seen from the autocorrelation function of the Scargle periodogram in Fig.\,\ref{autocorroffourier1}. The two highest peaks are situated at $\delta f = 0.03835081\,\mathrm{d}^{-1}$ and $\Delta f = 0.31799938\,\mathrm{d}^{-1}$, and we call these values the small spacing (the separation of the peak groups) and large spacing (the separation of the high amplitude regions) from here on. The latter also appears as a significant frequency in the Scargle periodogram at $0.318087\pm0.000016\,\mathrm{d}^{-1}$ ($221\pm7\,\mathrm{ppm}$), while the former one is responsible for the overall global beating pattern visible in the light curve.

\begin{figure}
\resizebox{\hsize}{!}{\includegraphics{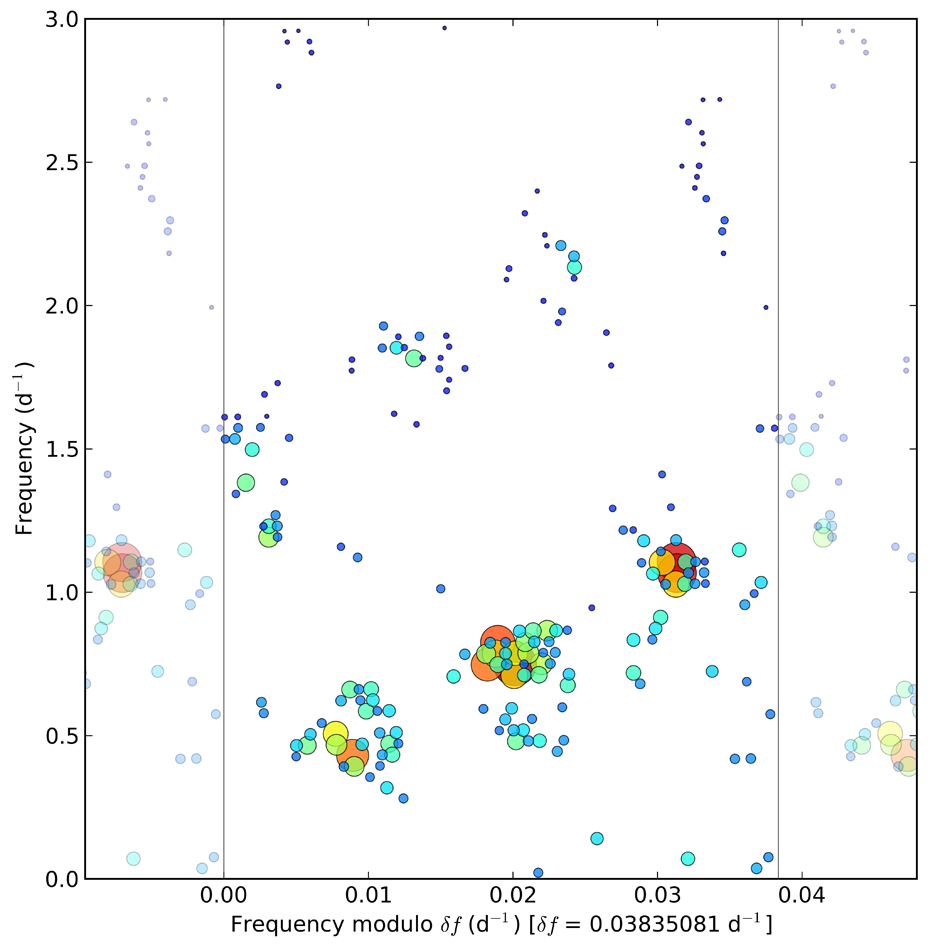}} 
\caption{\'{E}chelle diagram of the selected peaks in the frequency analysis of KIC\,4931738 shown between 0 and $3\,\mathrm{d}^{-1}$ calculated using $\delta f$. Size and colour of the symbols are connected with the amplitude of the peaks (larger symbol and colour from blue to red is higher amplitude).}
\label{echelle1a}
\end{figure}

\begin{figure}
\resizebox{\hsize}{!}{\includegraphics{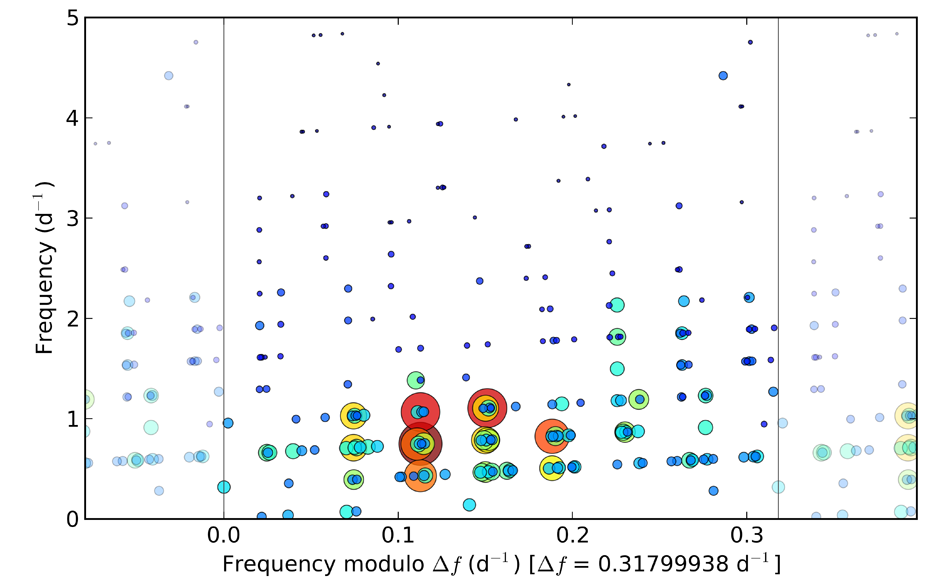}} 
\caption{\'{E}chelle diagram of the selected peaks in the frequency analysis of KIC\,4931738 shown between 0 and $5\,\mathrm{d}^{-1}$ calculated using $\Delta f$, plotted in a similar manner to Fig.\,\ref{echelle1a}.}
\label{echelle1b}
\end{figure}

We detect further details from an analysis of the \'{e}chelle diagram of the 262 peaks which met our selection criterion. The \'{e}chelle diagram constructed with $\delta f$ (see Fig.\,\ref{echelle1a}) shows clear groupings of peaks, while the \'{e}chelle diagram constructed using $\Delta f$ (see Fig.\,\ref{echelle1b}) shows well defined vertical ridges. The groupings in Fig.\,\ref{echelle1a} represent the different high amplitude regions of the Scargle periodogram. Although the different regions overlap slightly in frequency (the $y$ axis in Fig.\,\ref{echelle1a}), they are still well separated when we look at the corresponding modulo values (the $x$ axis). This means that what we see is not one long series of equidistantly spaced peak groups from 0.3 to $2.3\,\mathrm{d}^{-1}$, but six separate regions separated by $\Delta f$. Within these regions, peaks in dense groups which have an equal spacing of $\delta f$ occur. 

From Fig.\,\ref{echelle1b} it is clear that the latter three regions are slightly shifted relative to the first three, as the modulo value for the highest peaks in the first three regions is $\sim0.113\,\mathrm{d}^{-1}$, while it is $\sim0.226\,\mathrm{d}^{-1}$ for the last three. Thus the first three regions are separated by $\Delta f+0.112833\,\mathrm{d}^{-1}$ from the second three. Given that $2\times\Delta f+0.112833\,\mathrm{d}^{-1}=0.748832\,\mathrm{d}^{-1}$ equals the frequency of the dominant peak in the Scargle periodogram (the difference is $0.000026\,\mathrm{d}^{-1}$, while the Rayleigh-limit of the data set is $1/T = 0.000868$), we interpret the peaks in the three regions above $1.25\,\mathrm{d}^{-1}$ as low order combinations of peaks which can be found in the three regions below $1.25\,\mathrm{d}^{-1}$. This is confirmed by an automated search for linear combinations using a selected subset of the significant frequencies following \citet{2012AN....333.1053P}. We restricted the search to second or third order combinations formed by two individual frequencies. Using the same amplitude criterion and frequency precision as in \citet{2012AN....333.1053P}, looking at the 46 peaks which have $\mathrm{SNR} > 10$ in a $1\,\mathrm{d}^{-1}$ window (see Table\,\ref{combinations1}), we find none of the strong peaks below $1.2\,\mathrm{d}^{-1}$ to be combinations, while almost all above that limit are low order combinations.

For clarity purposes, we have restricted ourselves to the frequency range from 0.3 to $2.3\,\mathrm{d}^{-1}$ during the explanation of the observed pattern. However, further examination of Fig.\,\ref{echelle1a} and Fig.\,\ref{echelle1b} reveals that there are at least two more peak regions visible above $2.3\,\mathrm{d}^{-1}$, fitting perfectly into the described structure, representing higher order combinations with even lower amplitudes.

\begin{table}
\caption{Fourier parameters (frequencies ($f_j$), amplitudes ($A_j$), and possible linear combination identifications) of peaks having a signal-to-noise ratio (SNR) above 10 when computed in a $1\,\mathrm{d}^{-1}$ window after prewhitening for KIC\,4931738.}
\label{combinations1}
\renewcommand{\arraystretch}{1.00}
\centering
\begin{tabular}{l r r c}
\hline\hline
ID & $f\,(\mathrm{d}^{-1})$ & $A\,(\mathrm{ppm})$ & Note\\
\hline
$f_1    $&$ 0.748806 $&$ 4845.9 $&$13f_\mathrm{SB2} - 2f_\mathrm{rot}$  \\
$f_2    $&$ 1.105147 $&$ 3788.7 $&$18f_\mathrm{SB2} - 2f_\mathrm{rot}$  \\
$f_3    $&$ 1.066815 $&$ 3701.1 $&$14f_\mathrm{SB2} + 1f_\mathrm{rot}$  \\
$f_4    $&$ 0.824316 $&$ 2747.6 $&$14f_\mathrm{SB2} - 2f_\mathrm{rot}$  \\
$f_5    $&$ 0.430768 $&$ 2309.7 $&$ 5f_\mathrm{SB2} + 1f_\mathrm{rot}$  \\
$f_6    $&$ 0.746897 $&$ 2377.1 $&$13f_\mathrm{SB2} - 2f_\mathrm{rot}$  \\
$f_7    $&$ 0.785833 $&$ 1520.1 $&$10f_\mathrm{SB2} + 1f_\mathrm{rot}$  \\
$f_8    $&$ 0.710397 $&$ 1447.8 $&$ 9f_\mathrm{SB2} + 1f_\mathrm{rot}$  \\
$f_9    $&$ 1.028388 $&$ 1400.7 $&$17f_\mathrm{SB2} - 2f_\mathrm{rot}$  \\
$f_{10} $&$ 1.104149 $&$ 1373.8 $&$18f_\mathrm{SB2} - 2f_\mathrm{rot}$  \\
$f_{11} $&$ 0.787182 $&$ 1177.0 $&$10f_\mathrm{SB2} + 1f_\mathrm{rot}$  \\
$f_{12} $&$ 0.506300 $&$ 1222.1 $&$ 6f_\mathrm{SB2} + 1f_\mathrm{rot}$  \\
$f_{13} $&$ 0.866086 $&$  756.8 $&  \\
$f_{14} $&$ 0.750643 $&$  874.3 $&$13f_\mathrm{SB2} - 2f_\mathrm{rot}$  \\
$f_{15} $&$ 0.788076 $&$  797.2 $&$10f_\mathrm{SB2} + 1f_\mathrm{rot}$  \\
$f_{16} $&$ 0.468003 $&$  780.5 $&$ 9f_\mathrm{SB2} - 2f_\mathrm{rot}$  \\
$f_{17} $&$ 1.191988 $&$  701.7 $&  \\
$f_{18} $&$ 0.392532 $&$  686.8 $&$ 8f_\mathrm{SB2} - 2f_\mathrm{rot}$  \\
$f_{19} $&$ 0.826217 $&$  631.2 $&$14f_\mathrm{SB2} - 2f_\mathrm{rot}$  \\
$f_{20} $&$ 0.712112 $&$  443.6 $&$ 9f_\mathrm{SB2} + 1f_\mathrm{rot}$  \\
$f_{21} $&$ 0.785159 $&$  635.0 $&$10f_\mathrm{SB2} + 1f_\mathrm{rot}$  \\
$f_{22} $&$ 1.382161 $&$  499.7 $&$22f_\mathrm{SB2} - 2f_\mathrm{rot}$  \\
$f_{23} $&$ 0.465986 $&$  527.3 $&$ 9f_\mathrm{SB2} - 2f_\mathrm{rot}$  \\
$f_{24} $&$ 1.815646 $&$  468.7 $&$ f_1 + f_3 $ \\
$f_{25} $&$ 0.480420 $&$  478.3 $&  \\
$f_{26} $&$ 0.471656 $&$  469.7 $&$ 9f_\mathrm{SB2} - 2f_\mathrm{rot}$  \\
$f_{27} $&$ 1.105772 $&$  403.0 $&$18f_\mathrm{SB2} - 2f_\mathrm{rot}$  \\
$f_{28} $&$ 1.230342 $&$  328.2 $&$ f_{14} + f_{25} $  \\
$f_{29} $&$ 2.133550 $&$  305.4 $&$ f_2 + f_9 $ \\
$f_{30} $&$ 1.497647 $&$  286.1 $&$ f_3 + f_5 $ \\
$f_{31} $&$ 1.147825 $&$  285.1 $&  \\
$f_{32} $&$ 1.852782 $&$  233.8 $&$ f_1 + f_{10} $ \\
$f_{33} $&$ 2.171871 $&$  153.2 $&$ f_2 + f_3 $ \\
$f_{34} $&$ 1.534805 $&$  151.9 $&$ f_1 + f_7 $ \\
$f_{35} $&$ 2.209311 $&$  132.9 $&$ f_2 + f_{10} $ \\
$f_{36} $&$ 1.892722 $&$   86.0 $&$ f_2 + f_{11} $ \\
$f_{37} $&$ 4.420396 $&$   81.6 $&$ 3f_2 + f_{10} $ \\
$f_{38} $&$ 1.928587 $&$   81.1 $&$ f_4 + f_{10} $ \\
$f_{39} $&$ 2.297335 $&$   57.9 $&$ f_2 + f_{17} $ \\
$f_{40} $&$ 2.258822 $&$   56.9 $&$ f_3 + f_{17} $ \\
$f_{41} $&$ 3.123140 $&$   36.6 $&$ f_3 + 2f_9 $ \\
$f_{42} $&$ 3.238734 $&$   27.2 $&$ f_2 + 2f_3 $ \\
$f_{43} $&$ 3.940221 $&$   19.5 $&$ 2f_{10} + 2f_{13} $ \\
$f_{44} $&$ 4.754004 $&$   12.3 $&$ 3f_{10} + 3f_{25} $ \\
$f_{45} $&$12.868686 $&$    7.5 $&  \\
$f_{46} $&$19.256448 $&$    6.8 $&  \\
\hline
\end{tabular}
\tablefoot{For a complete table of frequencies and their errors see Table\,\ref{frequtable1}, while there is further explanation given in the text.}
\end{table}

These observations suggest that we need to interpret the signal of the first three regions only. We would like to point out, that the observed structure of power in the Fourier domain, such as consecutive high amplitude modes with an equal spacing of $\delta f$, forming a long continuous series except for two phase shifts (in frequency) between the consecutive higher amplitude regions which have a spacing of $\Delta f$, is very similar to the structure -- even of the amplitudes -- of the interference pattern of two closely spaced frequencies.

\subsubsection{Seismic interpretation}\label{seismicint1}

\begin{figure}
\resizebox{\hsize}{!}{\includegraphics{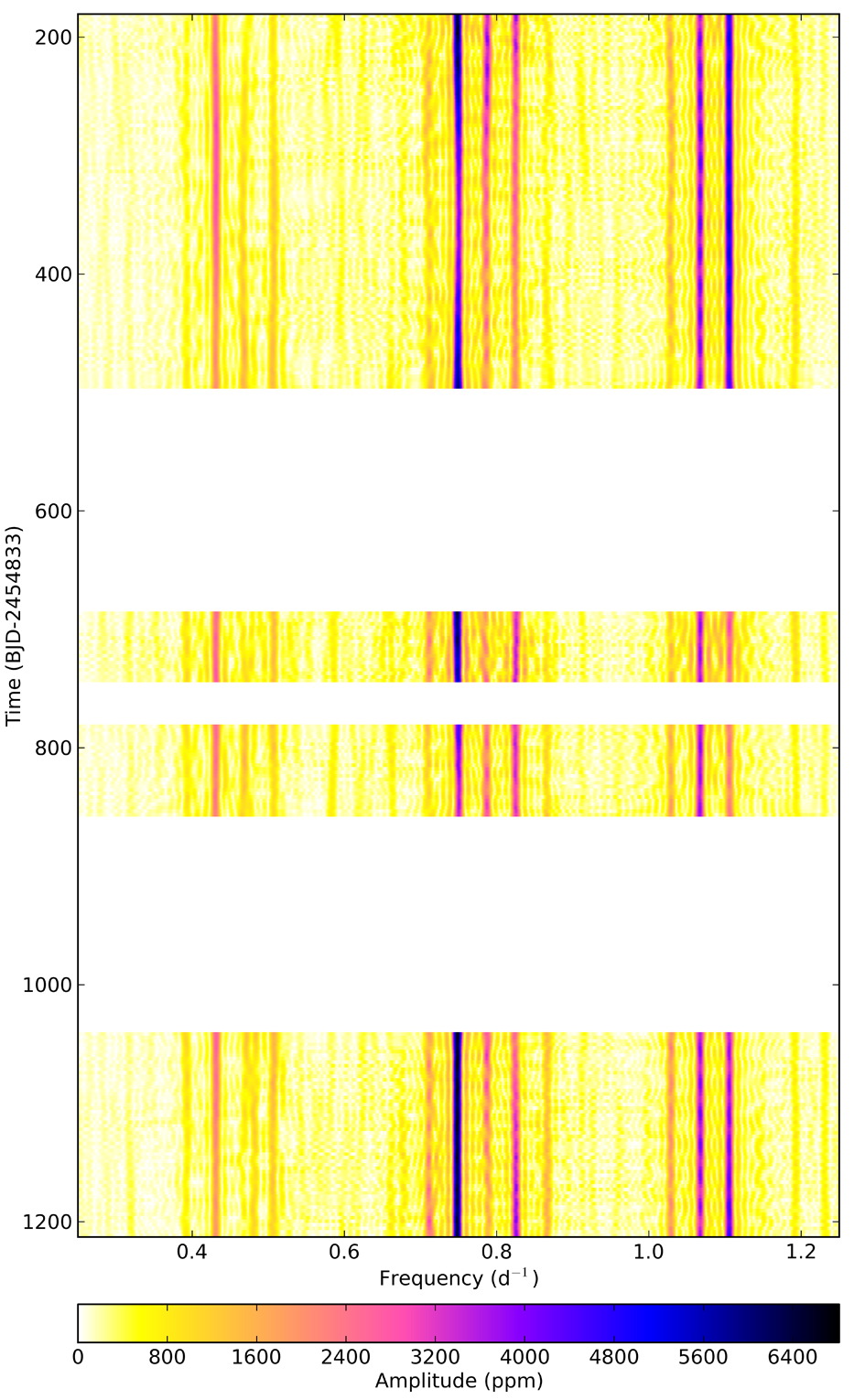}} 
\caption{Short-time Fourier transformation of the region of dominant oscillation modes (from 0.25 to $1.25\,\mathrm{d}^{-1}$) in the amplitude spectrum (window width of $120\,\mathrm{d}$), white voids correspond for those intervals where the duty cycle was less than $80\%$.}
\label{stft1}
\end{figure}

It is common practice to obtain the short-time Fourier transform (STFT) of the light curve to perform a time-frequency analysis of the data set \citep[see, e.g.,][]{2010_degroote_phd}. This is done by calculating the discrete Fourier transform (DFT) of consecutive windowed subsets of the data (we used rectangular windows to maximise the frequency resolution) over the full timebase of the observations, and comparing these consecutive DFTs to each other to see if frequencies, amplitudes, or phases show temporal evolution. The STFT in Fig.\,\ref{stft1} shows that all frequencies are stable over time, which would be atypical for any kind of stochastic excitation mechanism. Also, although all large amplitude peak regions show a very dense fine structure (e.g., $f_1-f_6\approx0.0019\,\mathrm{d}^{-1}$, slightly above the Rayleigh limit, but not meeting the \citeauthor{1978Ap&SS..56..285L} criterion), comparing the frequency analysis of the first half of the data set to the one of the second half, we see that even these structures are stable over time. This leaves us with two options: either the oscillations are excited by tidal forces (which is suggested by the structure in the periodogram along with the relatively close orbit and significant eccentricity), or internally by the $\kappa$-mechanism, which is the basic mechanism in pulsating B-type stars on the main sequence \citep[$\beta$ Cep and SPB stars, see, e.g.,][]{1999AcA....49..119P}. Independently of the excitation mechanism, the observed frequencies are in the range of $g$ modes.

\begin{figure}
\resizebox{\hsize}{!}{\includegraphics{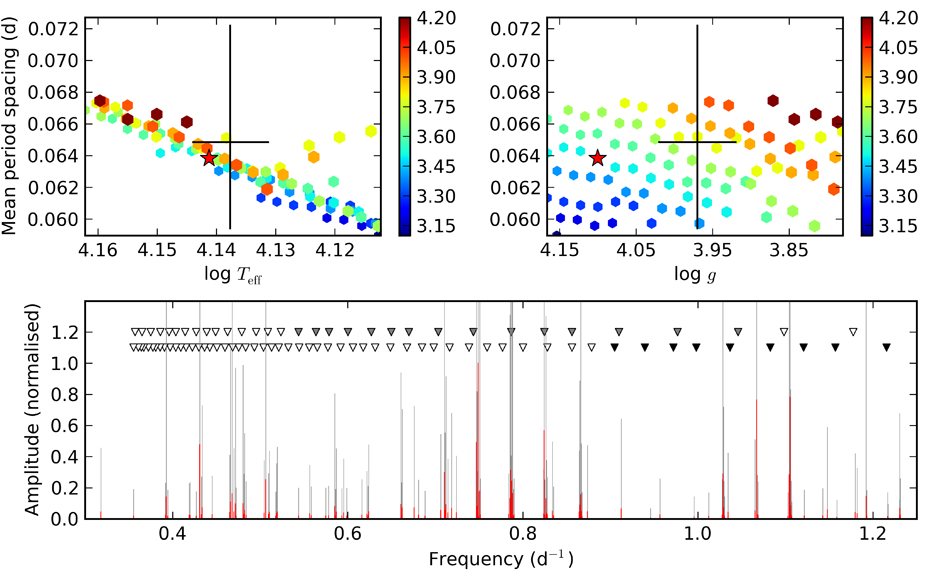}} 
\caption{The mean period spacing of $l=1$ modes (averaged over a large range of radial orders) from models ($Z = 0.01$, $\alpha_{\mathrm{ov}}=0.2$, $X=0.7$) within $4\sigma$ of the derived fundamental parameters compared to the spacing values of the 4 strongest modes of KIC\,4931738 around $0.75\,\mathrm{d}^{-1}$ plotted against $T_\mathrm{eff}$ (\textit{upper left}) and $\log g$ (\textit{upper right}). The scatter in median spacing values around the diagonal in the first panel is caused by the shape of the period spacings (the number and depth of dips in the $\Delta p(p)$ function). A black cross marks the observed mean spacing and the $T_\mathrm{eff}$ and $\log g$ value determined from spectroscopy, with the extents denoting the spread in the measured spacing values and the estimated error in the observed fundamental parameters. Colour represents the mass in units of solar mass. (\textit{lower panel} ) The normalised selected peaks from the periodogram of KIC\,4931738 are plotted in a similar manner than in Fig.\,\ref{fourier1}, with the theoretical frequencies from a selected model (which is marked with a red asterisk on the upper panels) marked by triangles: $l=1$ modes are plotted in the top row, while $l=2$ modes are plotted in the bottom, and frequencies expected to be excited are marked by filled symbols.}
\label{pulsationcheck1}
\end{figure}

In SPB stars one expects that heat-driven $g$ modes of the same low degree and of consecutive high radial order are equidistant in period space. We made a compatibility check to see if models having similar parameters to the primary component would show spacings similar to the measured ones. For this test we used the 4 strongest peaks around $0.75\,\mathrm{d}^{-1}$, and compared their period spacings to the mean spacing values of $l=1$ and $l=2$ modes \citep{2001A&A...366..166D} from models \citep{2008Ap&SS.316...83S}. We found (see Fig.\,\ref{pulsationcheck1}) that the mean period spacing of $l=1$ modes is compatible with the measured spacing of the 4 peaks in question. However, we cannot explain the period spacing of the high amplitude peaks around $0.43\,\mathrm{d}^{-1}$ and $1.07\,\mathrm{d}^{-1}$, which have a value a factor of $\sim3$ and $\sim0.5$, while having the same, equal spacing in frequency. 

\begin{figure}
\resizebox{\hsize}{!}{\includegraphics{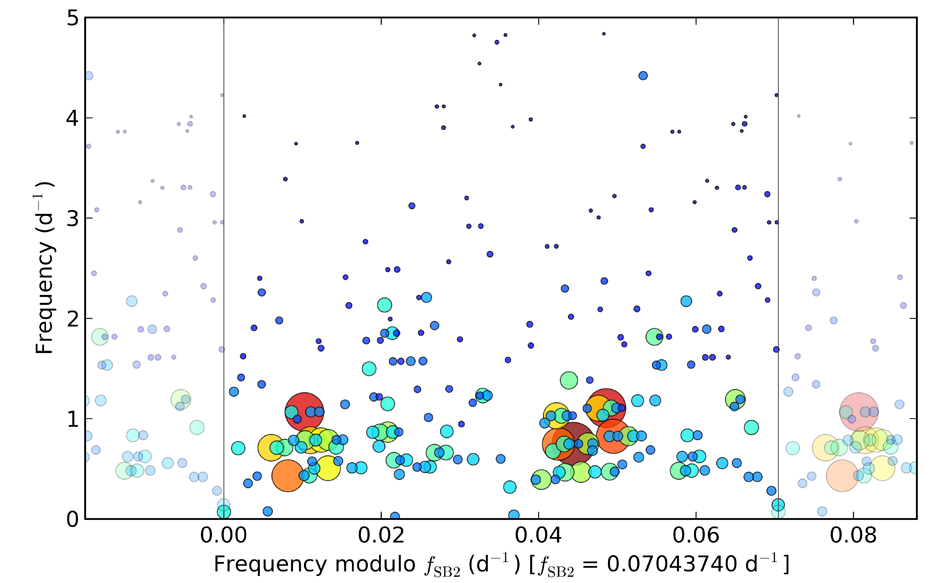}} 
\caption{\'{E}chelle diagram of the selected peaks in the frequency analysis of KIC\,4931738 shown between 0 and $5\,\mathrm{d}^{-1}$ calculated using $f_\mathrm{SB2}$, plotted in a similar manner to Fig.\,\ref{echelle1a}.}
\label{echelle1c}
\end{figure}

Depending on the orbital parameters and the eigenfrequencies of the components, pulsation modes can be excited by tidal forces. In such case the forcing frequencies of the tide-generating potential, \[f_\mathrm{T} = kf_\mathrm{SB2} + mf_\mathrm{rot},\] with $k,m\in\mathbb{Z}$, must come into resonance with the free oscillation frequencies of the components. Thus we expect to see structure in the \'{e}chelle diagram calculated using $f_\mathrm{SB2}$. Indeed, this is the case here. Fig.\,\ref{echelle1c} shows two clear (but not exactly sharp) ridges for the strongest peaks in the three large amplitude regions. The lowest possible $k,m$ combinations are the most probable, so we tried to find an optimal rotation period, which provides a frequency pattern matching the observed one.

The two simplest choices to reproduce the frequency pattern occur for $m=1,-2$ combinations and $f_\mathrm{rot} = (1/6)f_\mathrm{SB2}$ or $f_\mathrm{rot} = (7/6)f_\mathrm{SB2}$. The first value of $f_\mathrm{rot}$ is incompatible with the measured projected rotational velocity $v \sin i_\mathrm{rot} = v_\mathrm{min} = 10.5\pm1.0\,\mathrm{km\,s}^{-1}$, as it would require a star with a radius of at least $17.7\,R_{\sun}$, which is significantly higher than what is possible for stars in this $\log g - T_\mathrm{eff}$ range. The second value is compatible with the $v \sin i_\mathrm{rot}$ measurement, as such rotation frequency would set the radius at $2.53\,R_{\sun}/\sin i_\mathrm{rot}$. As a rough comparison, models displayed in Fig.\,\ref{pulsationcheck1}, have a median radius of $3.16\,R_{\sun}$, which would translate into an $i_\mathrm{rot}$ of $\sim53\degr$. These models have a median mass of $3.7\,\mathcal{M}_{\sun}$, which, using the $\mathcal{M}\sin^3 i_\mathrm{orb}$ value from Table\,\ref{fundparams1}, gives an $i_\mathrm{orb}$ of $\sim62\degr$. The fact that $i_\mathrm{rot}$ and $i_\mathrm{orb}$ are close estimates, illustrates that the suggested $f_\mathrm{rot} = (7/6)f_\mathrm{SB2}$ is compatible with both the observations and the model. Using $f_\mathrm{rot} = (7/6)f_\mathrm{SB2}$ and $m=1,-2$ we can reasonably match almost all peaks in Table\,\ref{combinations1} which were not identified as linear combinations before in Sect.\,\ref{spacings1}. The observed peaks have an average absolute deviation of $0.002669\,\mathrm{d}^{-1}$ from the exact tidal frequencies. This is $\sim3.3$ times lower than what is expected ($(1/8)f_\mathrm{SB2}$) for a random distribution of frequencies and the suggested model, and only 3 times higher than the Rayleigh frequency. Those very few peaks, which cannot be matched in Table\,\ref{combinations1} are interpreted as excited free oscillation modes of the primary. We note here, that we see no sign of rotational modulation in the light curve.

There are a couple of weak peaks at higher frequencies (see Table\,\ref{frequtable1}), which cannot be explained or matched using linear combinations, thus we interpret them as low amplitude $p$ modes. 

\subsection{KIC\,6352430}\label{frequencynotes2}
\begin{figure*}
\resizebox{\hsize}{!}{\includegraphics{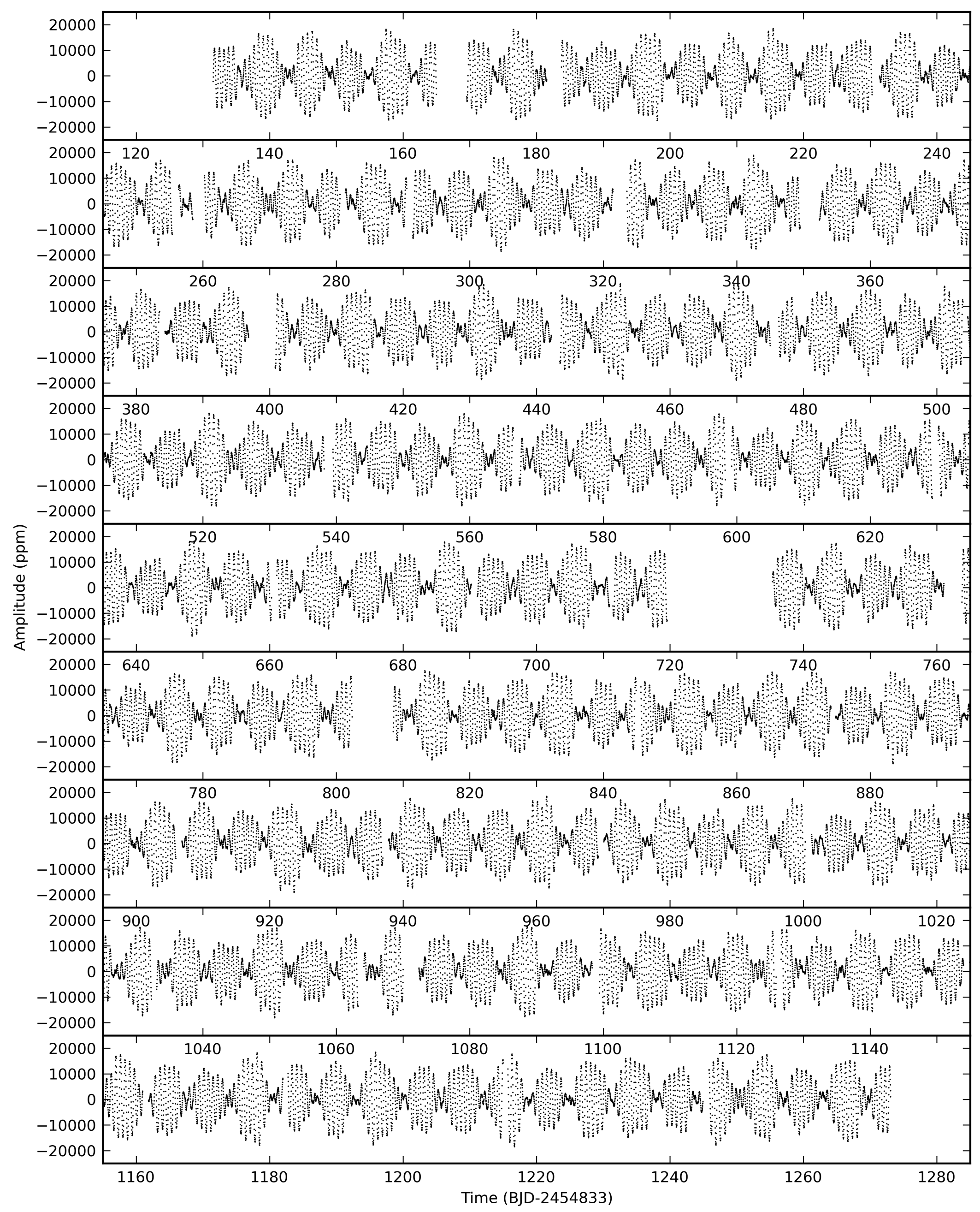}} 
\caption{The full reduced \textit{Kepler} light curve (black dots, brightest at the top) of KIC\,6352430.}
\label{lightcurve2}
\end{figure*}

The cleaned and detrended \textit{Kepler} light curve (see Fig.\,\ref{lightcurve2}) containing 51903 data points covers 1141.54 days starting from BJD 2454964.51190674 with a 92.91\% duty cycle.

\begin{figure*}
\resizebox{\hsize}{!}{\includegraphics{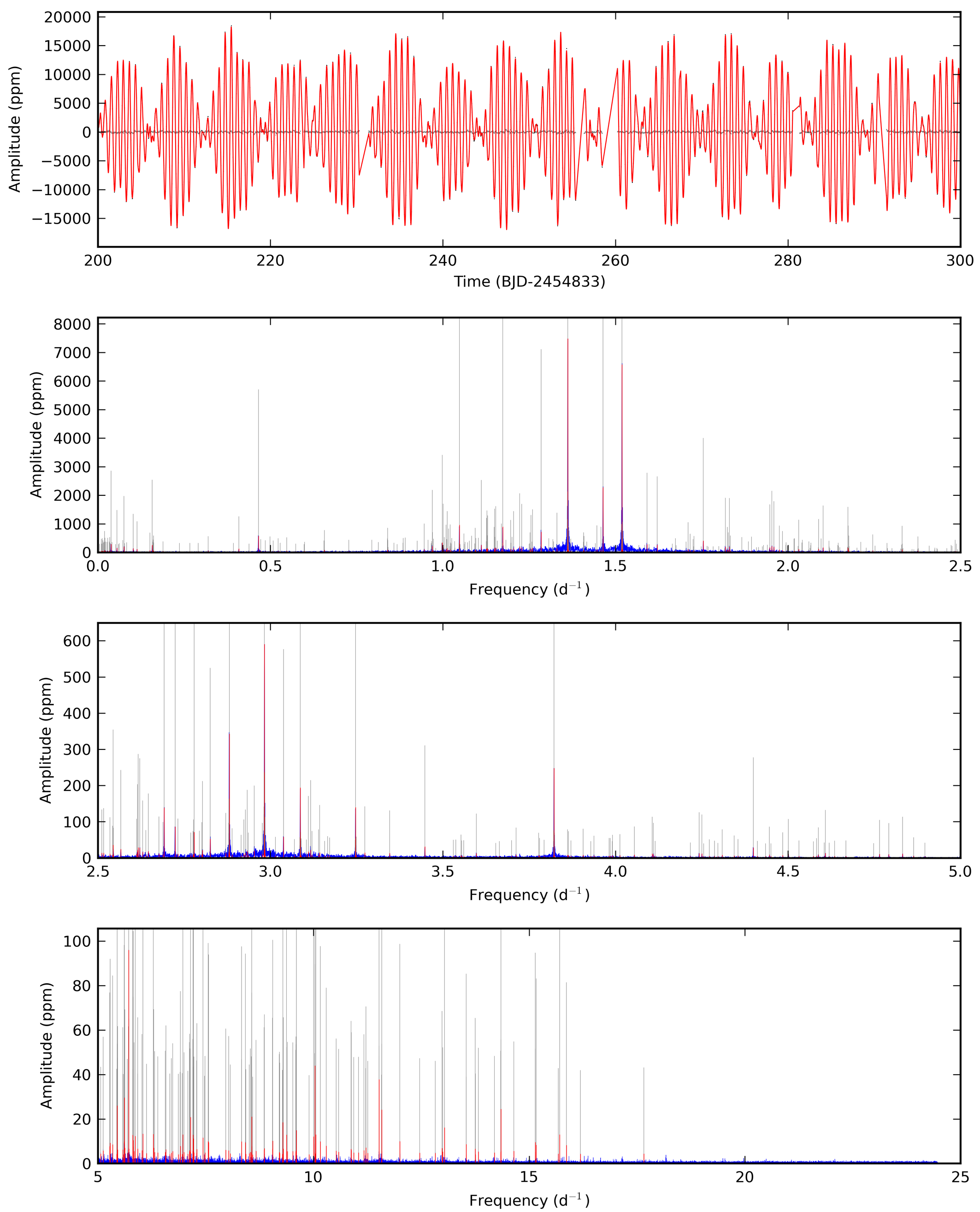}} 
\caption{(\textit{upper panel}) A zoom-in section of the reduced \textit{Kepler} light curve (black dots, brightest at the top) and residuals (gray dots) of KIC\,6352430 after prewhitening with a model (red solid line) constructed using the set of 1175 significant frequencies -- see text for further explanation. (\textit{lower three panels}) The Scargle periodogram of the full \textit{Kepler} light curve (blue solid line) showing the 555 selected frequencies (red vertical lines). For better visibility, the red lines are repeated in gray in the background, after multiplying their amplitude with a factor ten, and the signal from outside the plotted ranges is prewhitened for each panel.}
\label{fourier2}
\end{figure*}

The iterative prewhitening procedure returned 1784 frequencies of which 1175 met our significance criteria. The Scargle periodogram is shown in Fig.\,\ref{fourier2}. The model constructed using this set of frequencies provides a variance reduction of 99.97\% while bringing down the average signal levels from 216.6--217.0--4.5--3.6--3.2 ppm to 1.5--1.4--0.6--0.5--0.2 ppm, measured in $2\,\mathrm{d}^{-1}$ windows centred around 1, 2, 5, 10, and $20\,\mathrm{d}^{-1}$, respectively. Most of the power is distributed between 0.25 and $2.5\,\mathrm{d}^{-1}$: 785 peaks contribute to a total power density of $4.83\times10^7\,\mathrm{ppm}^2\mathrm{d}$, while there are 107 peaks between 2.5 and $5\,\mathrm{d}^{-1}$ ($2.55\times10^5\,\mathrm{ppm}^2\mathrm{d}$), and 149 peaks above $5\,\mathrm{d}^{-1}$ ($1.20\times10^3\,\mathrm{ppm}^2\mathrm{d}$). The 134 peaks below $0.25\,\mathrm{d}^{-1}$ provide a power density of $1.25\times10^6\,\mathrm{ppm}^2\mathrm{d}$, and the statement on the low frequencies in Sect.\,\ref{frequencynotes1} is valid here as well, i.e., one part of this power comes from small and unavoidable instrumental trends left in the data, while the other part has a physical origin, and most probably related to granulation noise and/or low amplitude pulsations. 555 of the significant frequencies met the selection criterion and are listed in Table\,\ref{frequtable2}.

The binary orbital period of $26.551\pm0.019$ days ($f_\mathrm{SB2} = 0.037663\pm0.000027\,\mathrm{d}^{-1}$) can also be found in the frequency analysis of the light curve at $0.037698\pm0.000010\,\mathrm{d}^{-1}$ with an amplitude of $285\pm6\,\mathrm{ppm}$, along with four of its harmonics with amplitudes of $197\pm5\,\mathrm{ppm}$, $108\pm4\,\mathrm{ppm}$, $69\pm3\,\mathrm{ppm}$, and $38\pm2\,\mathrm{ppm}$, for $2f_\mathrm{SB2}$, $3f_\mathrm{SB2}$, $4f_\mathrm{SB2}$, and $5f_\mathrm{SB2}$, respectively. The appearance of the harmonics is a sign of a slight ellipsoidal modulation. These are some of the most significant frequencies below $0.25\,\mathrm{d}^{-1}$. There is no sign of eclipses in the light curve phased with the orbital period. The light variation due to oscillations is too complicated to be prewhitened with the aim to interpret the photometric variability due to the orbital motion alone.

\subsubsection{Spacings in Fourier space}

The Fourier spectrum is dominated by two very strong and four medium peaks between 1 and $1.6\,\mathrm{d}^{-1}$. The frequency difference of the two dominant peaks creates the $\sim6.8$ day beating pattern which dominates the light curve. There is another group of peaks clearly standing out from the rest in amplitude between 2.6 and $3.1\,\mathrm{d}^{-1}$, these are mostly low order combinations of the strongest peaks between 1 and $1.6\,\mathrm{d}^{-1}$. Negative combinations (where -- in the simplest case -- the frequency of the combination peak is not the sum, but the difference of the two parent peaks) of the three strongest modes occur below $0.25\,\mathrm{d}^{-1}$.

The clear presence of both positive and negative combinations is also shown by an automatic combination frequency search, similar to the one we have used in Sect.\,\ref{spacings1}, but now looking at the 29 peaks having an $\mathrm{SNR} > 20$ (see Table\,\ref{combinations2}), and using a slightly different set of parameters to, e.g., allow for negative combinations. The use of a higher signal-to-noise cutoff -- which was required to reach a similar limited sample as in Sect.\,\ref{spacings1} -- was needed because of the different distribution of the frequencies and amplitudes of the most significant peaks. This test not only confirms the presence of various combination frequencies, but also shows that there are several independent modes present in higher frequency regions, unlike in the frequency spectrum of KIC\,4931738.

\begin{table}
\caption{Fourier parameters (frequencies ($f_j$), amplitudes ($A_j$), and possible linear combination identifications) of peaks having a signal-to-noise ratio (SNR) above 20 when computed in a $1\,\mathrm{d}^{-1}$ window after prewhitening for KIC\,6352430.}
\label{combinations2}
\centering
\begin{tabular}{l r r c}
\hline\hline
ID & $f\,(\mathrm{d}^{-1})$ & $A\,(\mathrm{ppm})$ & Note\\
\hline
$f_1   $&$  1.361690 $&$ 7478.7 $& \\
$f_2   $&$  1.518726 $&$ 6574.6 $& \\
$f_3   $&$  1.463729 $&$ 2271.3 $& \\
$f_4   $&$  1.047486 $&$  959.5 $&$ 3f_1 - 2f_2 $ \\
$f_5   $&$  1.173030 $&$  893.9 $& \\
$f_6   $&$  1.284234 $&$  711.4 $& \\
$f_7   $&$  2.982459 $&$  590.3 $&$ f_2 + f_3 $ \\
$f_8   $&$  0.465004 $&$  570.2 $& \\
$f_9   $&$  1.754439 $&$  400.1 $&$ 2f_3 - f_5 $ \\
$f_{10}$&$  0.997712 $&$  341.2 $& \\
$f_{11}$&$  2.880422 $&$  342.5 $&$ f_1 + f_2 $ \\
$f_{12}$&$  0.037698 $&$  285.4 $&$ f_\mathrm{SB2} $ \\
$f_{13}$&$  0.157032 $&$  253.8 $&$ f_2 - f_1 $ \\
$f_{14}$&$  3.821662 $&$  247.3 $& \\
$f_{15}$&$  3.086319 $&$  193.7 $& \\
$f_{16}$&$  3.246498 $&$  139.5 $& \\
$f_{17}$&$  2.691767 $&$  138.6 $&$ f_2 + f_5 $ \\
$f_{18}$&$  5.713156 $&$   95.9 $&$ 3f_6 + 4f_8 $ \\
$f_{19}$&$  2.723390 $&$   86.0 $&$ 2f_1 $ \\
$f_{20}$&$  3.037473 $&$   57.6 $&$ 2f_2 $ \\
$f_{21}$&$ 10.027681 $&$   43.9 $& \\
$f_{22}$&$ 11.515490 $&$   37.7 $& \\
$f_{23}$&$  5.611949 $&$   29.5 $& \\
$f_{24}$&$  5.446646 $&$   26.0 $&$ 4f_1 $ \\
$f_{25}$&$ 14.340933 $&$   24.4 $&$ 2f_{10} + 4f_{15} $ \\
$f_{26}$&$ 11.576811 $&$   24.1 $& \\
$f_{27}$&$  8.562751 $&$   20.9 $& \\
$f_{28}$&$ 13.034207 $&$   16.0 $&$ f_2 + f_{22} $ \\
$f_{29}$&$ 15.702644 $&$   12.9 $& \\
\hline
\end{tabular}
\tablefoot{For a complete table of frequencies and their errors see Table\,\ref{frequtable2}.}
\end{table}

\begin{figure}
\resizebox{\hsize}{!}{\includegraphics{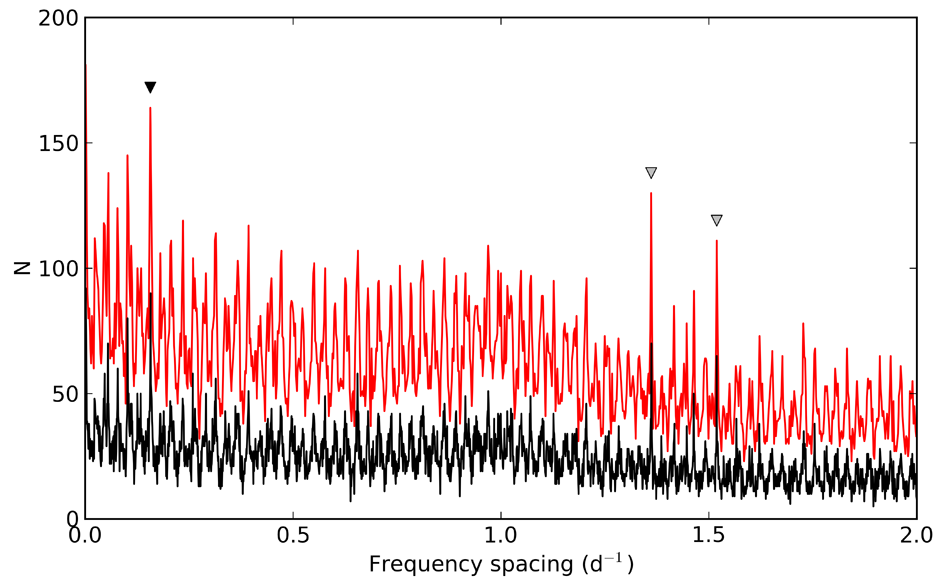}} 
\caption{Histogram of the distribution of frequency pairs between the 368 selected peaks in the frequency analysis of KIC\,6352430 situated between 0 and $4\,\mathrm{d}^{-1}$. The histogram with a bin width of $2.5/T$ is plotted with red, while the histogram with a bin width of $1/T$ is plotted with a black solid line. The spacing $\delta_1 f$ is marked with a black triangle, while $\Delta_1 f$ and $\Delta_2 f$ are marked with grey triangles.}
\label{freqpairshistogram2a}
\end{figure}

\begin{figure}
\resizebox{\hsize}{!}{\includegraphics{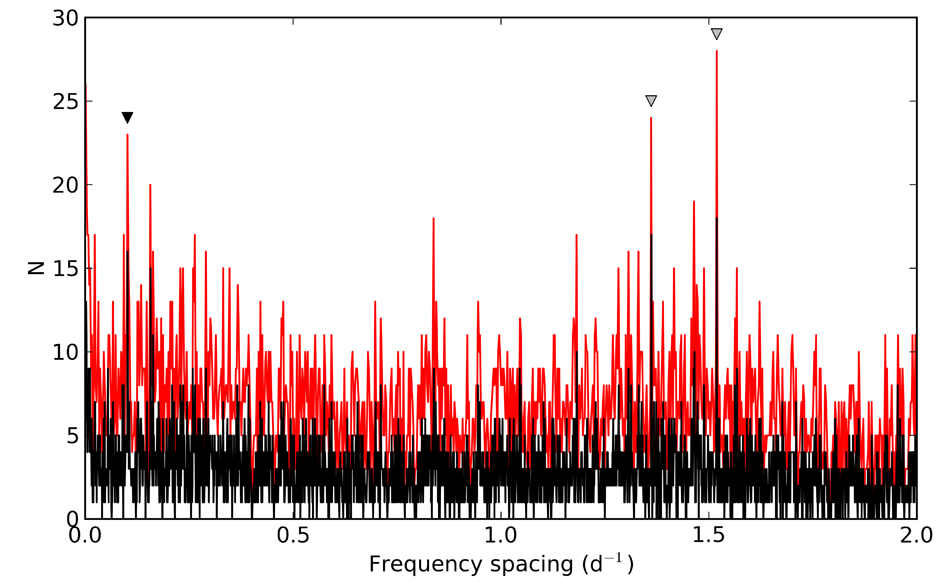}} 
\caption{Histogram of the distribution of frequency pairs between the 187 selected peaks in the frequency analysis of KIC\,6352430 situated above $4\,\mathrm{d}^{-1}$. The histogram with a bin width of $2.5/T$ is plotted with red, while the histogram with a bin width of $1/T$ is plotted with a black solid line. The spacing $\delta_2 f$ is marked with a black triangle, while $\Delta_1 f$ and $\Delta_2 f$ are marked with grey triangles.}
\label{freqpairshistogram2b}
\end{figure}

Although it is not as strikingly clear as for KIC\,4931738, the periodogram of KIC\,6352430 also shows spacings in frequency. As the autocorrelation function is heavily dependent on the amplitudes, the presence of two strong peaks requires a different approach than for KIC\,4931738. We checked the distribution of frequency differences for all possible frequency pairs in both the low and high frequency region (see Fig.\,\ref{freqpairshistogram2a} and Fig.\,\ref{freqpairshistogram2b}). This method returns two small spacing values, $\delta_1 f = 0.157\pm0.001\,\mathrm{d}^{-1}$ (more dominant in the lower frequency region), and $\delta_2 f = 0.102\pm0.001\,\mathrm{d}^{-1}$ (more dominant towards higher frequencies), and two large spacing values, $\Delta_1 f = 1.361\pm0.001\,\mathrm{d}^{-1}$, and $\Delta_2 f = 1.519\pm0.001\,\mathrm{d}^{-1}$, both visible in the full frequency range, but dominant in the high frequency region. These spacing values seem to be connected to frequency values and differences, as $\delta_1 f = f_2-f_1 = 0.157086\,\mathrm{d}^{-1}$, $\delta_2 f = f_3-f_1 = 0.102088\,\mathrm{d}^{-1}$, $\Delta_1 f = f_1 = 1.361640\,\mathrm{d}^{-1}$, and $\Delta_2 f = f_2 = 1.518726\,\mathrm{d}^{-1}$.

\begin{figure}
\resizebox{\hsize}{!}{\includegraphics{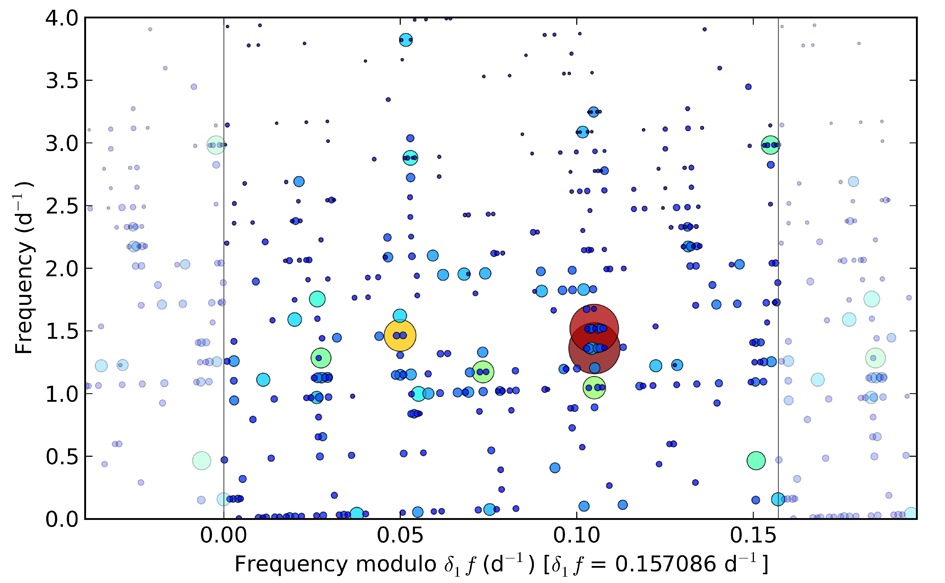}} 
\caption{\'{E}chelle diagram of the selected peaks in the frequency analysis of KIC\,6352430 shown between 0 and $4\,\mathrm{d}^{-1}$ calculated using $\delta_1 f$, plotted in a similar manner to Fig.\,\ref{echelle1a}.}
\label{echelle2a}
\end{figure}

\begin{figure}
\resizebox{\hsize}{!}{\includegraphics{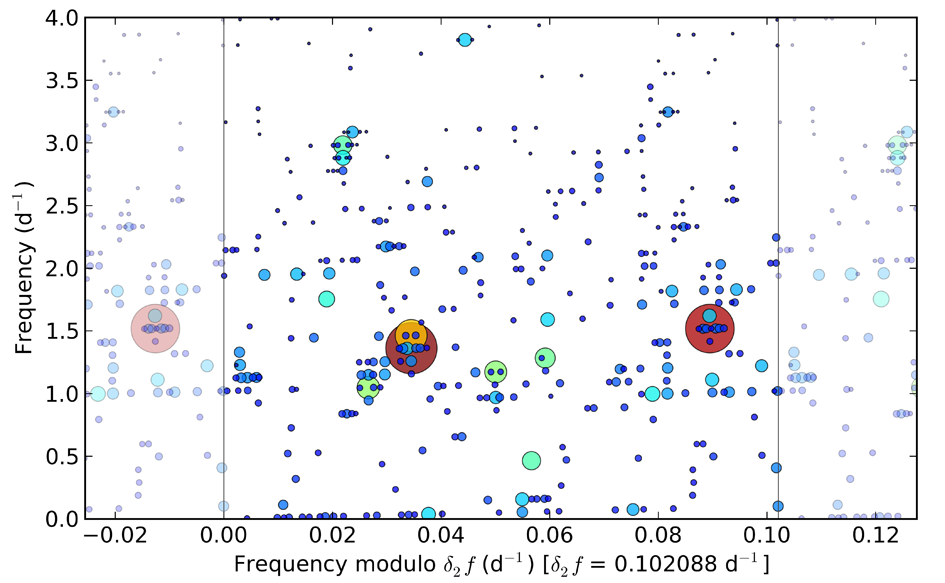}} 
\caption{\'{E}chelle diagram of the selected peaks in the frequency analysis of KIC\,6352430 shown between 0 and $4\,\mathrm{d}^{-1}$ calculated using $\delta_2 f$, plotted in a similar manner to Fig.\,\ref{echelle1a}.}
\label{echelle2b}
\end{figure}

\begin{figure*}
\resizebox{\hsize}{!}{\includegraphics{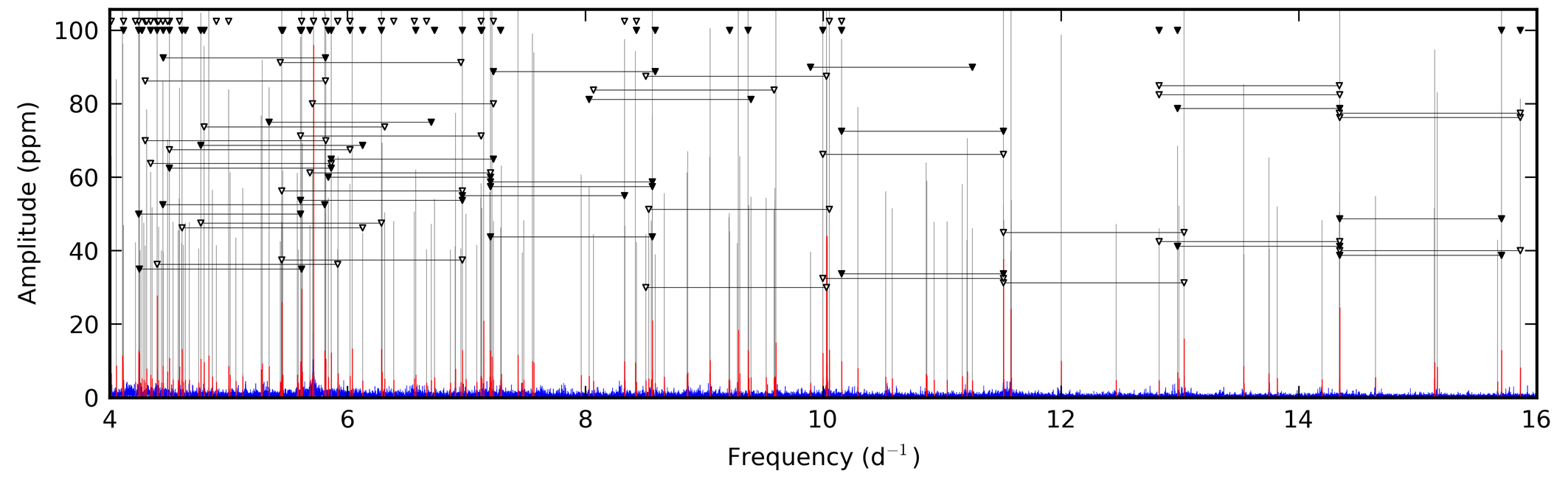}} 
\caption{Frequency pairs among the 187 selected peaks in the frequency analysis of KIC\,6352430 situated above $4\,\mathrm{d}^{-1}$. Pairs with $\delta_1 f$ are plotted with black triangles placed at 100 ppm, pairs with $\delta_2 f$ are plotted with empty triangles placed at 102.5 ppm, pairs with $\Delta_1 f$ are plotted with connected black triangles, and pairs with $\Delta_2 f$ are plotted with connected empty triangles. For the key to other elements we refer to Fig.\,\ref{fourier2}.}
\label{freqpairs2b}
\end{figure*}

Looking at the \'{e}chelle diagram constructed with $\delta_1 f$ (see Fig.\,\ref{echelle2a}) or $\delta_2 f$ (see Fig.\,\ref{echelle2b}) of the 555 peaks which met our selection criterion, we can see vertical ridges of not only two, but several equidistantly spaced frequencies: $\Delta_1 f$ and $\Delta_2 f$ dominate in the pure $p$ mode regime, which looks similar to, but more pronounced than in the case of HD\,50230 \citep{2012A&A..542A..88D}. The frequency pairs in Fig.\,\ref{freqpairs2b} show a clear pattern in the region above $6\,\mathrm{d}^{-1}$, where the equally spaced frequencies seem to create a quasi continuous series of peaks.

\subsubsection{Seismic interpretation}

The \'{e}chelle diagram calculated using $f_\mathrm{SB2}$ shows no structure at all, which suggests that the oscillations have no connection to the binary nature of the star. Typical for heat-driven oscillations in B stars, all modes are stable over time. Not surprisingly, we see $g$ mode oscillations in the low frequency regime (characteristic of SPB stars), but the high frequency region is remarkable. We observe pulsation signal with frequencies and frequency spacings typical for $p$ modes, but which are not expected to be excited by theory in this relatively low temperature range near $13\,000\,\mathrm{K}$, and were not seen in any of the well analysed B stars. We can exclude that this signal is coming from the secondary, as -- beyond the big difference in luminosity -- the frequency spacings are connected to strong pulsation modes seen in the $g$ mode region of the primary.

We interpret the signal in the $p$ mode region as a result of nonlinear resonant mode excitation due to the high-amplitude $g$ modes. This explanation is suggested by the fact that the frequency spacing values in this region match the two highest amplitude $g$ modes, and also by the richness of the frequency spectrum in low order combination frequencies. This could explain the appearance of all four observed spacing values. A compatibility check similar to the one we have done in Sect.\,\ref{seismicint1} showed that the excited peaks fall in the low frequency end of the region where eigenfrequencies of low order low degree $p$ modes are situated, and that the frequency spacings are compatible with the observed ones.

Similar nonlinear effects are expected to be seen in large-amplitude nonradial oscillators \citep[e.g.,][]{1997A&A...321..159B}, such as white dwarfs \citep[e.g.,][]{2008PASP..120.1043F}, $\delta$\,Sct stars \citep[e.g.,][]{2005A&A...435..955B}, and $\beta$\,Cep variables \citep[e.g.,][]{2009A&A...506..111D,2009A&A...506..269B}, but excitation of modes in the $p$ mode regime through resonant coupling with dominant $g$ modes were, as far as we are aware, not yet observed before in SPBs.

%%%%%%%%%%%%%%%%%%
%%%Conclusions %%%
%%%%%%%%%%%%%%%%%%

\section{Conclusions}\label{conclusions}
Our observational study is the first detailed asteroseismic analysis of main sequence B-type stars by means of more than three years of \textit{Kepler} space photometry and high-resolution ground based spectroscopy. It led to the classification and detailed description of two SB2 binary systems. The derived observational binary and seismic constraints provide suitable starting points for in-depth seismic modelling of the two massive primary stars.

The first system, KIC\,4931738 is a binary consisting of a B6\,V primary and a B8.5\,V secondary, both slow rotators on the main sequence, with an orbital period of $14.197\pm0.002$\,d and an eccentricity of $0.191\pm0.001$. The observed pulsation spectrum is consistent with tidally excited $g$ modes in the primary component. Tidal excitation of pulsation modes was observed only in a few unevolved main-sequence binaries before, e.g., by \citet{2000A&A...359..539D}, \citet{2002A&A...384..441W}, \citet{2002MNRAS.333..262H}, \citet{2009A&A...508.1375M}, and \citet{2011ApJS..197....4W}.

The second system, KIC\,6352430 is multiple system with two observed components, a moderate rotator main sequence B7\,V primary and a slow rotator F2.5\,V secondary contributing less than 10\% of the total light, with an orbital period of $26.551\pm0.019$\,d and an eccentricity of $0.370\pm0.003$. The primary star shows typical $g$ mode SPB pulsations, but in addition we detect a rich spectrum of modes in the $p$ mode region, atypical for such a low temperature SPB star. As these modes are not expected to be excited by the $\kappa$ mechanism at this temperature, we explain their presence by nonlinear resonant excitation by the two dominant $g$ modes, as supported by the observed high number of combination frequencies exhibiting four main spacing values, which are all connected to these $g$-mode frequencies. Such excitation was not observed before in any SPB star.

In our next paper we will analyse the 6 single B-type pulsators from the sample described in Sect.\,\ref{selection} and followed-up be \textit{Kepler} through our \textit{Kepler} GO programme.

%%%%%%%%%%%%%%%%%%%%%%
%%%Acknowledgements%%%
%%%%%%%%%%%%%%%%%%%%%%

\begin{acknowledgements}
The research leading to these results has received funding from the European Research Council under the European Community's Seventh Framework Programme (FP7/2007--2013)/ERC grant agreement n$^\circ$227224 (PROSPERITY), as well as from the Belgian Science Policy Office (Belspo, C90309: CoRoT Data Exploitation). KU acknowledges financial support by the Spanish National Plan of R\&D for 2010, project AYA2010-17803. We thank the spanish Night-Time Allocation Committee (CAT) for awarding time to the proposals 82-NOT4/12A and 61-Mercator3/11B. This paper includes data collected by the \textit{Kepler} mission. Funding for the \textit{Kepler} mission is provided by the NASA Science Mission directorate. Some of the data presented in this paper were obtained from the Multimission Archive at the Space Telescope Science Institute (MAST). STScI is operated by the Association of Universities for Research in Astronomy, Inc., under NASA contract NAS5-26555. Support for MAST for non-HST data is provided by the NASA Office of Space Science via grant NNX09AF08G and by other grants and contracts. This work is partly based on observations made with the William Herschel Telescope, which is operated on the island of La Palma by the Isaac Newton Group in the Spanish Observatorio del Roque de los Muchachos (ORM) of the Instituto de Astrof\'{i}sica de Canarias (IAC).
\end{acknowledgements}

\bibliographystyle{aa}
\bibliography{KeplerBstarsI}

%%%%%%%%%%%%%%%%%%%%%%%%%%
%%%Appendix for on-line%%%
%%%%%%%%%%%%%%%%%%%%%%%%%%

\Online
\begin{appendix}
\section{Tables}\label{tables}

\begin{longtable}{c c c c c c c c c}
\caption{\label{frequtable1} Fourier parameters (frequencies ($f_j$), amplitudes ($A_j$), and phases ($\theta_j$)) of the selected peaks having a signal-to-noise ratio (SNR) above 4 for KIC\,4931738 when computed in a $1\,\mathrm{d}^{-1}$ window after prewhitening. The displayed SNR values are calculated in a window of $1\,\mathrm{d}^{-1}$ or $3\,\mathrm{d}^{-1}$ centred on the given frequency, or from the full periodogram (from $0\,\mathrm{d}^{-1}$ to $24.47\,\mathrm{d}^{-1}$). A complete list of all significant frequencies is available upon request.}\\
\hline\hline
$f\,(\mathrm{d}^{-1})$ & $\epsilon_f\,(\mathrm{d}^{-1})$ & $A\,(\mathrm{ppm})$ & $\epsilon_A\,(\mathrm{ppm})$ & $\theta\,(2\pi/\mathrm{rad})$ & $\epsilon_{\theta}\,(2\pi/\mathrm{rad})$ & \multicolumn{3}{c}{SNR ($1\,\mathrm{d}^{-1}$, $3\,\mathrm{d}^{-1}$, full)}\\
\hline
\endfirsthead
\caption{continued.}\\
\hline\hline
$f\,(\mathrm{d}^{-1})$ & $\epsilon_f\,(\mathrm{d}^{-1})$ & $A\,(\mathrm{ppm})$ & $\epsilon_A\,(\mathrm{ppm})$ & $\theta\,(2\pi/\mathrm{rad})$ & $\epsilon_{\theta}\,(2\pi/\mathrm{rad})$ & \multicolumn{3}{c}{SNR ($1\,\mathrm{d}^{-1}$, $3\,\mathrm{d}^{-1}$, full)}\\
\hline
\endhead
\hline
\endfoot
  0.021751 &   0.000023 &   100.7 &     4.9 &   0.3398 &   0.0483 &   4.3 &  13.4 &  40.2 \\
  0.036850 &   0.000018 &   148.0 &     5.8 &  -0.4875 &   0.0383 &   5.6 &  17.6 &  53.4 \\
  0.070457 &   0.000014 &   264.8 &     8.0 &  -0.0015 &   0.0301 &   7.3 &  24.5 &  73.8 \\
  0.076021 &   0.000023 &   107.3 &     5.1 &   0.3809 &   0.0477 &   4.4 &  13.7 &  41.4 \\
  0.140869 &   0.000016 &   215.9 &     7.1 &  -0.4818 &   0.0336 &   6.5 &  21.5 &  65.0 \\
  0.280883 &   0.000023 &    99.3 &     4.8 &  -0.4767 &   0.0482 &   4.4 &  13.3 &  40.0 \\
  0.318087 &   0.000016 &   221.0 &     7.2 &  -0.0511 &   0.0335 &   6.5 &  21.6 &  65.3 \\
  0.355267 &   0.000024 &    92.5 &     4.6 &   0.3168 &   0.0500 &   4.2 &  12.7 &  38.0 \\
\end{longtable}

\begin{longtable}{c c c c c c c c c}
\caption{\label{frequtable2} Fourier parameters (frequencies ($f_j$), amplitudes ($A_j$), and phases ($\theta_j$)) of the selected peaks having a signal-to-noise ratio (SNR) above 4 for KIC\,6352430 when computed in a $1\,\mathrm{d}^{-1}$ window after prewhitening. The displayed SNR values are calculated in a window of $1\,\mathrm{d}^{-1}$ or $3\,\mathrm{d}^{-1}$ centred on the given frequency, or from the full periodogram (from $0\,\mathrm{d}^{-1}$ to $24.47\,\mathrm{d}^{-1}$). A complete list of all significant frequencies is available upon request.}\\
\hline\hline
$f\,(\mathrm{d}^{-1})$ & $\epsilon_f\,(\mathrm{d}^{-1})$ & $A\,(\mathrm{ppm})$ & $\epsilon_A\,(\mathrm{ppm})$ & $\theta\,(2\pi/\mathrm{rad})$ & $\epsilon_{\theta}\,(2\pi/\mathrm{rad})$ & \multicolumn{3}{c}{SNR ($1\,\mathrm{d}^{-1}$, $3\,\mathrm{d}^{-1}$, full)}\\
\hline
\endfirsthead
\caption{continued.}\\
\hline\hline
$f\,(\mathrm{d}^{-1})$ & $\epsilon_f\,(\mathrm{d}^{-1})$ & $A\,(\mathrm{ppm})$ & $\epsilon_A\,(\mathrm{ppm})$ & $\theta\,(2\pi/\mathrm{rad})$ & $\epsilon_{\theta}\,(2\pi/\mathrm{rad})$ & \multicolumn{3}{c}{SNR ($1\,\mathrm{d}^{-1}$, $3\,\mathrm{d}^{-1}$, full)}\\
\hline
\endhead
\hline
\endfoot
  0.004284 &   0.000032 &    29.3 &     2.0 &   0.1039 &   0.0666 &   4.2 &   8.5 &  22.2 \\
  0.009853 &   0.000027 &    46.7 &     2.5 &  -0.3229 &   0.0551 &   5.1 &  10.8 &  29.2 \\
  0.011087 &   0.000032 &    30.1 &     2.0 &  -0.4449 &   0.0673 &   4.2 &   8.4 &  22.2 \\
  0.012010 &   0.000027 &    49.2 &     2.5 &  -0.4390 &   0.0552 &   5.1 &  10.8 &  29.3 \\
  0.014463 &   0.000028 &    41.8 &     2.4 &   0.3818 &   0.0574 &   5.0 &  10.4 &  28.0 \\
  0.015752 &   0.000031 &    34.4 &     2.2 &   0.1716 &   0.0647 &   4.3 &   8.9 &  23.6 \\
  0.017720 &   0.000029 &    36.3 &     2.2 &   0.1253 &   0.0608 &   4.6 &   9.5 &  25.5 \\
  0.018686 &   0.000020 &    74.7 &     3.1 &   0.2021 &   0.0419 &   7.5 &  15.1 &  42.3 \\
\end{longtable}

\end{appendix}

\end{document}